\documentclass[12pt]{article}
\usepackage[latin1]{inputenc}
\usepackage{latexsym}
\usepackage{graphicx}
\usepackage{epsfig}
\usepackage{tikz-feynman}
\usepackage{verbatim}
\usepackage{subfigure}

\def\*{\`}

\def\.{\cdot}
\def\la{\lambda}
\def\s{\sigma}

\def\o{\over}
\def\v{\vec}
\def\a{\alpha}

\def\c{\gamma}
\def\b{\beta}
\def\d{\delta}
\def\f{\phi}

\def\C{\Gamma}

\def\ep{\epsilon}
\def\p{\partial}

\def\+{\bigoplus}

\def\({\left(}
\def\){\right)}
\def\[{\left[}
\def\]{\right]}
\def\l.{\left.}
\def\r.{\right.}

\def\be{\begin{equation}}
\def\ee{\end{equation}}
\def\bea{\begin{eqnarray}}
\def\eea{\end{eqnarray}}
\def\nn{\nonumber \\ &&}

\def\acca{\right.\nn\left.}
\def\ber{\begin{array}}
\def\eer{\end{array}}

\newcommand{\ks}{k\hspace{-0.9ex}/}

\newcommand{\vs}{v\hspace{-0.9ex}/}

\begin{document}

\begin{titlepage}

\noindent
Octobers, 2022\hfill

\vskip 1in
\begin{center}
\def\thefootnote{\fnsymbol{footnote}}
{\large \bf The heavy tetra-quark states after the discovery of tetra-charm states.}

\vskip 0.3in

C. Becchi \footnote{E-Mail: becchi@ge.infn.it}

\vskip .2in

     {\em Dipartimento di Fisica, Universit\`a di Genova\\
          and\\ Istituto Nazionale di Fisica Nucleare, Sezione di 
Genova\\
              via Dodecaneso 33, I-16146, Genoa, Italy}

\end{center}

\vskip .4in

\begin{abstract}

This paper is devoted to the study of exclusive charm and bottom bound states which are suitable, at least partially, for a non-relativistic description. Starting from a simple choice of quarkonia and tetra-quark wave functions we study their production cross sections, total widths and more interesting decay properties.

In particular, we compute the production cross sections of the ground state scalar quarkonia getting results compatible with the values appearing in the literature. We also compute the production cross sections of the lower energy states of the charm and bottom tetra-quarks, their decay widths and the decay rates in two $\mu\bar\mu$ pairs and in the $q\bar q\mu\bar\mu$ channel.

Our results on the tetra-charm are consistent with the data published by the LHCb Collaboration and those on the tetra-bottom show a reasonable agreement with the indications published by the CMS Collaboration. Our cross section analyses are based on interpolations based on the gluon-gluon luminosity grids published by the NNPDF and CTEQ Collaborations. The hard amplitude calculations are based on QCD in the tree approximation. We think that a careful improvement of our method and a refinement of the experimental data after the foreseen luminosity increase at the LHC will make possible valuable tests of QCD at the intermediate energies.
\end{abstract}

\vfill

\setcounter{footnote}{0}

\end{titlepage}

\section{Introduction}

The discovery of heavy quarkonia states $c\bar c$, $b\bar b$ and $c\bar b$ and the more recent detection of fully charmed
tetra-quarks has revived the interest in non-relativistic quark physics since it turns out that the radius of the heavy quarkonia is larger than the quark Compton wavelength.

A natural consequence of this is the interest in fully heavy tetra-quark states. We shall denote by the capital letter $Q$ the heavy Quarks. The physics of these compounds is expected to give important QCD tests because Quarks can be considered fixed sources and the binding forces could be easily identified in lattice QCD also accounting for the many body forces \cite{12}. For  recent review articles see \cite{nb}.

In practice these considerations should concern the Quarks with masses of some $GeV$.
Strictly speaking this weakly applies to the  charm flavor while it is well satisfied by the bottom one. Top Quarks do not live long enough to feel inter quark binding forces.

We shall consider charmonium and bottomium states together with tetra-charm and tetra-bottom with the purpose of treating the whole multi-Quark physics in a coherent and consistent framework based on Quantum Field Theory in the non-relativistic limit. Quarkonia will give us interesting tests of our non-relativistic approach to the dynamics of both flavors that we shall also apply to tetra-Quarks. 
Clearly the first problem one must face with is the existence of bound or resonant tetra-Quark states. There are theoretical papers, partly based on lattice QCD, discussing this problem \cite{1}\cite{11}. The role of many body forces is however rarely discussed \cite{12}. 

It is in general considered that tetra-quarks might be, either compact bound states, or  molecules of quarkonia, i.e. be made of pairs of quarkonia bound by hadronic interactions analogous to those binding nuclei. If this  were the nature of all the detected tetra-quarks the interest in their existence would be strongly diminished. On the contrary compact bound states might give important tests of QCD.
Considering the physical differences between the two kind of binding mechanism let us remark that  the radii of the molecular  tetra-Quark  states should be substantially larger than the sum of those of the Quarkonia, while that of the compact bounds is expected to be smaller. As a consequence the  production cross section should be at least by two orders of magnitude larger than that in the compact case. The compactness choice has been already assumed in many papers among which  \cite{12},  \cite{1}  \cite{5}, and  \cite{6}. The results of  the present paper are in strong agreement with this choice. Of course, the compactness of the  fully heavy tetra-Quarks does not imply anything about the bound states containing light quarks.

Recently the discussion about the existence of tetra-Quarks has been intensified by the  results published by the LHCb Collaboration \cite{4} which, studying the production of double $\mu\bar\mu$ pairs, has found some evidence of tetra-charm resonances. In particular they have detected strong indications of resonances decaying into $Y(1S)\mu\bar\mu$ appearing over a background generated by the $\mu\bar\mu$ decay of pairs of $J/\psi$. We shall discuss this point in greated detail in the following.

Obviously, the existence of tetra-charm resonances should imply that of tetra-bottom ones because the short-distance Coulomb like forces act more strongly on the bottom Quarks. 

Following the line of our former papers \cite{5} and \cite{6} we continue the analysis of the possibility of detecting the production of the tetra-bottom at the LHC.
On account of \cite{1} we shall assume in the non-relativistic approximation multi-Gaussian wave functions for both Quarkonia and tetra-Quarks of charm and bottom flavors. We shall denote by ${\cal Q}$ the lower energy vector Quarkonia and by $\eta_Q$ the pseudoscalar ones. Furthermore we shall denote by ${\cal T}_Q$ the tetra-Quark states. For Quarkonia we shall choose the orbital wave function
\be \Psi_{\cal Q}(\v r_1-\v r_2, \ \v P)= {1\o 2^{3\o2} \pi^{9\o4}D^{3\o2}_{\cal Q}} e^{-{(\v r_1-\v r_2)\o 2 D^{2}_{\cal Q}}+i\v P\.{\v r_1+\v r_2\o2}}\ .\label{1}\ee

For the ${\cal T}_Q$ states, following \cite{1}, we choose the product of three Gaussian factors, the first two factors depending respectively on the distance between the Quarks and that between the anti-Quarks. The third factor depends on the distance between the two Q and the two anti-Q centers of mass. Taking into account the charge conjugation invariance we set
\be\Psi_{\cal T}(\v r_1,\v r_2 ,\v r_3,\v r_4, \v P)={1\o2^{3\o2} \pi^{15\o4}d^3\d^{3\o2}}e^{-{ (\v r_1+\v r_2-\v r_3-\v r_4)^2\o8\d^2 }-{(\v r_1-\v r_2)^2+(\v r_3-\v r_4)^2\o 2 d^2}}e^{i\v P\. {\v r_1+\v r_2+\v r_3+\v r_4\o4}}\ .\label{2}\ee Clearly this choice of the wave functions is based on the non-relativistic character of the heavy Quark states. 

We shall compute in Section 2 the radii $D_{\cal Q}$ of Quarkonia from their decay rate into $\mu\bar\mu $ pairs.

In Section 3 we shall evaluate the paramenters $d$ and $\d$ using the variational method.

In Section 4 
we shall recall the properties of the lower energy states of tetra-Quarks that we have
already discussed in references \cite{5} and \cite{6}. We shall represent these states in the second quantized form.

Once   chosen the wave functions  we can compute the more interesting decay processes of the Quarkonia and tetra-Quarks using the effective Hamiltonians which are computed in the Appendices applying Quantum Field Theory in the semi-classical (tree) approximation and in the non-relativistic framework for the heavy Quarks which is discussed in Appendix A.

A further important subject in the physics of the heavy-Quark compound states is their production cross section in the proton-proton collisions. 
In the parton model colliding protons appear as showers of partons, i.e. quarks, anti-quarks and gluons, whose momenta are essentially collinear\footnote{In order to evaluate the corrections due to non-vanishing transverse parton momenta $\v p_t$ we must know the mean square value $\overline{p_t^2}$ of $\v p_t$ . Assuming a normal  transverse momentum distribution  we can get an estimate of $\overline{p_t^2}$ of the produced $\eta_c$ at the LHCb from references \cite{setac} which give the $\eta_c$ production cross section without any transverse momentum 
selection and that for transverse momentum $p_t>6.5\ GeV$, the first cross section being $\s(0)\simeq25.4\pm 7.7\ \mu b$ while the second one amounts to $\s(6.5)\simeq1.26\pm .33\ \mu b$. From these data we can evaluate the mean square value of the $\eta_c$ transverse momenta getting $\overline{p_t^2}\sim 14\ GeV^2\sim 1.5\ m^2_{\eta_c}$. Other experiments concerning the $p_t$ distributions  give  $\overline{p_t^2}\sim 40\ GeV^2\sim .4\ m^2_{\Upsilon}$  for the $\Upsilon$ \cite{s1y} and $\overline{p_t^2}\sim 10^3\ GeV^2\sim .4\ m^2_{Z}$  for the intermediate boson $Z$ \cite{tz}. The most obvious correction  due to 
$p_t$ is the Lorents contraction of the cross section. This corresponds to a correction of roughly $-\overline{p_t^2}/(m^2(1+\exp(2y)))$, that is to less than $-12\ \%$ for the $\eta_c$ at the LHCb and less than $-9\ \%$ for the $\eta_b$ at the CMS apparatuses. These are of the order of the relativistic corrections.}, 
 \footnote{We  systematically sum   statistical and systematic errors.}.  The production of a  heavy particle can be seen as the fusion of two partons, one from each proton, creating the heavy particle. This is the Drell-Yan mechanism which is clearly described in reference \cite{8}.

In Section 5 we recall how the Drell-Yan production cross section of a particle of mass $M$  is related to the product of the modulus square of the transition amplitude between the two partons and the heavy particle and the luminosity \cite{10} of the crossing parton showers.

Based on the results of Section 5 and of Appendices B and D    in Section 5.1 we present the calculation   in a generic heavy flavor case of the ratio
 ${\cal R}$ of  the resonant production of two $\mu\bar\mu$ pairs to that due to the $\mu\bar\mu$ decay of two Quarkonia ${\cal Q}$.

In Section 5.2 we   compare and discuss  the values of the gluon-gluon  luminosities   giving the values we shall use in our cross section calculations. In particular, considering the published results,  we discuss the values of the gluon-gluon luminosities at the scales $ 3\ GeV$ and $6\ GeV$ for the LHCb apparatus and $10\ GeV$ and $20\ GeV$ for the CMS apparatus.

In Section  6, using  our evaluations of the transition amplitudes $ \eta\leftrightarrow 2 g $,  ${\cal T}\leftrightarrow 2 g $ in Appendix D.1 and  ${\cal T} \leftrightarrow q\bar q $ in Appendix D.2,
we compute  the production cross sections of $\eta_Q$'s and tetra-Quarks. We also compute  the  rates of the most important decay processes among which those into two $\mu\bar\mu$ pairs of the tetra-Quarks   are the easiest to compute and the most interesting from the experimental point of view. 
Together with the production cross section of tetra-Quarks 
we  compute   in both flavor cases the value of the above mentioned ratio ${\cal R}$. 

We shall show that   the detection of tetra-bottom resonances looks very difficult due to the small values of the production cross sections of  bottom  compound states.

On the contrary the resonant production of $Q\bar Q\mu\bar\mu$ that we study in Appendix B.2 could be identified using "b-tagging'' for $b\bar b$ pairs near their threshold. However, the efficiency of "$b \bar b$-tagging'' at about $1\ GeV$ above threshold seems to be very small   for transverse momenta of few $GeV$.

Beyond the electromagnetic decays we study the strong decays into gluons and light quarks. 
We shall also consider in Appendix C 
interesting channels such as the tetra-Quark decays into a $Q\bar Q$ pair together with two gluons or a light quark pair. From the lower order calculations, the branching ratios in these channels turn out to be the larger ones. Therefore, they give the most important contributions to the width of tetra-Quarks.

Particularly interesting is the calculation of the decay rates of the Quarkonia $\eta_Q$ into two gluons which   should give the larger contribution to the width of these scalar Quarkonia thus allowing a direct comparison with experimental results and hence a test of our method. The results, based on the values of $\a_S(m_Q^2)$ given by \cite{9}, are $\C_{\eta_c}\simeq 25\ MeV\ ,$ and $\C_{\eta_b}\simeq 4.6\ MeV$ which are consistent with the Particle Data Book data.

All these calculations shall be limited to the hard part of the amplitudes, that concerning high momentum transfers among the elementary constituents. What we miss is hadronization, however this is unavoidable. Consider for example the $\eta_c$ decays, there are more than fifteen final channels containing two, three and four, hadrons. Therefore, the quarks and gluons in the final states must be related to clusters of hadrons. Of course, hadronization modifies the rates but it should not change the order of magnitude of the results as it happens for the Quarkonia decay rates.

We shall systematically assign to the Quarkonia the mass $2m_Q$ and to the tetra-Quarks the mass $4m_Q$. We think that the errors due to these simplifying choices are negligible in the framework of the considered approximations.

\bigskip

\section{ ${\cal Q}\to \mu\bar\mu $ decays and Quarkonic radii.}

In order to determine the radii of Quarkonia $D$ appearing in the wave function given in Eq. (\ref{1}) we begin our analysis computing the rate of the decay ${\cal Q}\to \mu\bar\mu$  using the golden rule  \cite{10}.

Due to the Q annihilation nature of the processes, we shall represent the states in the Fock space and we shall use the effective Hamiltonian in the rest frame of the initial state. This Hamiltonian is given in Eq. (\ref{3}), in Appendix A,  taking into account that the momenta of the Quarks are negligible with respect of their masses. 
The initial state with momentum $\v P$ and $J_z=1$ is given by
\bea&&|{\cal Q}, 1 ,\v P >={1\o\sqrt3}\int d\v r_1d\v r_2 \Psi_{\cal Q}(\v r_1-\v r_2, \v P)\tilde A^\dag_{+,a}(\v r_1)\tilde B^\dag_{+,a}(\v r_2)|0>\nn={1\o\sqrt3\ 2^{3\o2}\pi^{9\o 4}D^{3\o2}}\int d\v r_1d\v r_2 e^{i\v P\.{\v r_1+\v r_2\o2}}e^{-{(\v r_1-\v r_2)^2\o 2 D^2}}\tilde A^\dag_{+,a}(\v r_1)\tilde B^\dag_{+,a}(\v r_2)|0>
\ ,
\label{5}\eea while the scalar Quarkonium state is 
\be|\eta_Q ,\v P >={1\o\sqrt3\ 4 \pi^{9\o 4}D^{3\o2}}\int d\v r_1d\v r_2 e^{i\v P\.{\v r_1+\v r_2\o2}}e^{-{(\v r_1-\v r_2)^2\o 2 D^2}}[A^\dag_{+,a}(\v r_1)\tilde B^\dag_{-,a}(\v r_2)-A^\dag_{-,a}(\v r_1)\tilde B^\dag_{+,a}(\v r_2)]|0>
\ .
\label{eta}\ee
Therefore in the rest frame of ${\cal Q}$ the transition amplitude is given by
$$ <0|b_{\la'}(\v q_2)a_{\la }(\v q_1)H_I|{\cal Q}, 1, \v 0>=-{ \a\ \zeta\o \sqrt3 \ 2^{5\o2}\pi^{5\o 4} m_Q^2\ D^{3\o2}\sqrt{q_1q_2}} \d(\v q_1+\v q_2)
(\bar u_\la(\v q_1) (\c_x+i\c_y)v_{\la'}(\v q_2)) \ ,$$ where $\zeta$ is the triple of the Quark electric charge, and the decay rate is given by
\bea&&\C_{{\cal Q} \to\mu\bar\mu}\sim { \a^2\ \zeta^2\o3\ 2^4\pi^{3\o2}\ m_Q^4\ D_\Upsilon^3} \int {d\v k\o k^2}\d(2(k-m_b))Tr(\ks(\c_x+i\c_y)\tilde\ks(\c_x-i\c_y)) \nn
={ 4\ \a^2\ \zeta^2
\o 9\ \sqrt\pi\ D ^3 \ m_Q^2}\simeq 0.25{\a^2\ \zeta^2\o \ m_b^2\ D^3}\ .\label{y2gamma}\eea Taking into account that in the
Particle Data Books one finds $\C_{J/\psi\to\mu\bar\mu}\sim 5.4\ keV$ and $\C_{\Upsilon \to\mu\bar\mu}\sim 1.3\ keV$ we can compute the radius of 
$\Upsilon$ finding 
\be D_\Upsilon\equiv D_b=\( 4\ \a ^2\o 9\sqrt\pi\ m_b^2 \C_{\Upsilon \to\mu\bar\mu}\)^{1\o3}\sim .8\ GeV^{-1}\ ,
\label{raggio} \ee
while for $J/\psi$ we have
\be D_c=\( 16\ \a ^2\o 9\sqrt\pi\ m_c^2 \C_{J/\psi \to\mu\bar\mu}\)^{1\o3}\sim 1.6\ GeV^{-1}\ .
\label{raggioc} \ee Now we can verify the validity of the non-relativistic approximation for both heavy flavors. Computing the product $m_Q^2 D_Q^2$ we find roughly 13 for the bottom and 6 for the charm. Therefore, we see that the non-relativistic approximation is more suitable for the bottom than for the charm but it should not be rejected even in the second case. 

\section{The wave function parameters.}
\bigskip

 Once assumed the form of the wave functions it is natural to compute its parameters by the variational method. However, this requires a good knowledge of the masses and of the potential energies. 
The values of the constituent masses of the heavy Quarks can be found in the literature where one finds the charm mass value $m_c= 1.55 \ GeV$ and the value of the bottom mass $m_b= 4.73 \ GeV\ .$

Concerning the potential energies, as discussed in \cite{12} the dual superconductor model of the QCD vacuum implies that the tetra-Quark binding forces at large distances are generated by a constant tension string similar in shape to the capital H letter connecting a pair of Quarks in color triplet state on one side to a pair of anti-Quarks also in color triplet state on the other side. According to the lattice calculations \cite{12} the value of the string tension should be about $\s=.16\ GeV^2$.

Assuming orbital $S$ waves as in Eq. (\ref{2}) both Quarks
and anti-Quarks must be in spin triplet states, thus there should exist nine lower energy states with total angular momentum ranging from zero to two and mass differences of the order of magnitude of $100\ MeV$. 

At short distances the binding energy is dominated by Coulomb like two body forces whose strength, the QCD strong coupling $\a_S(Q^2)$, depends on the momentum transfer and hence on the radius of the bound state \cite{10}.

However,  the effective value of $\a_S(1/D^2)$ at the considered momentum transfer values is quite uncertain. For this reason, we shall use the variational relation  for the Quarkonium  which relates the radius, the Quark mass, the string tension and the strong coupling (Equation \ref{var1})  to compute the strong coupling and then using the tetra-Quark variational relation we shall compute the wave function parameters in Equation (\ref{2}). Indeed from the relations 
\be {8\ \a_S\o 3\sqrt\pi}={3\o m_Q D_Q}-{2\s\o\sqrt\pi}D_Q^2\ .\label{var1}
\ee knowing the Quarkonium radii, i.e. $D_c=1.6\ GeV^{-1}$ for the charmonium and $D_b=.8\ GeV^{-1}$ for the bottomium one gets  the value of $\a_S$ appropriate for the two Quarkonia. We shall assume the same values for the corresponding tetra-Quarks because the average distances and hence the exchanged momenta are essentially the same for Quarkonia and tetra-Quarks. 

The same variational method allows the determination of the tetra-Quark wave function parameters $d$ and $\d$ identifying the minima in the positive quadrant of the function \footnote{ Notice that in the above function  the term corresponding to the string energy has been approximated replacing the average minimal length of the string by $d+ 1.6\ \d$. With regard to this choice notice that if the vectors joining the two (anti-)quarks, that we denote by $\v \rho$ and $\v\rho'$ and that joining the center of mass of the two pairs that we denote by $\v R$, are parallel the minimal string length is $R+(\rho+\rho')/2$, while if they are orthogonal the length is $R+(4-\sqrt3)(\rho+\rho')/2\ .$}
\be E(d\ , \d)\equiv{3\o 4\ m_Q }({1\o \d^2}+{4\o d^2})+{2\s\o\sqrt\pi}( d+ 1.6\ \d)-{8\ \a_S\o 3\sqrt\pi}\({1\o d}+\sqrt{2\o 2\ d^2+ \d^2}\)\ .\label{var2}
\ee 
It is apparent that   two equations identify uniquely the parameters $d$ and $\d$ as functions of the tension $\s$. In particular for the assumed value of the tension and $\a_S$ we find for the tetra-charm $\d_c=1.44 \ GeV^{-1}$ and 
$ d_c=1.87 \ GeV^{-1}$, while for the tetra-bottom we find $\d_b = .803\ GeV^{-1}$ and $ d_b=1.01 \ GeV^{-1}$.

Now we consider the tetra-Quarks.

\section{The states of the heavy tetra-Quarks ${\cal T}$}
\bigskip

Following the analysis in \cite{5} and \cite{6} we consider the  ground states of a non-relativistic $QQ\bar Q\bar Q$ system that we denote by ${\cal T}$ assuming   the wave function given in Eq. (\ref{2}).

QCD foresees that the most interacting states are those in which two anti-Quarks in color triplet, and hence with spin one, interact with two Quarks, in color anti-triplet and spin one. 
Thus we have nine possible states with total angular momentum 2, 1 and 0 and positive parity. The states with even angular momentum have positive charge conjugation while the others have negative charge conjugation. There is also a state of null angular momentum in which Quarks combine in color sextets and anti-Quarks in color anti-sextets. However, according to the QCD string model this state should be less interacting. 

The states with positive parity and charge conjugation should partially decay, even if with small branching ratio, in two $\mu\bar\mu$ pairs, as shown by the LHCb Collaboration. This seems to be the channel of maximum signal to background ratio and hence most suitable for detection provided that the cross section be sufficient. 
A detailed discussion of these states and of their decay channel is contained in \cite{5} and \cite{6}.

According to vector meson dominance the decay into two $\mu \bar\mu$ pairs is dominated by the two step tetra-Quark decay into a ${\cal Q}$ with emission of a $\mu\bar\mu$ pair followed by the ${\cal Q}$ decay into a further $\mu \bar\mu$ pair. The rate of the first decay step can be computed introducing the following Fock representation of the ${\cal T}$ based on the space wave function given in Eq. (\ref{2}). Understanding the sum over repeated color indices the tetra-Quark state with momentum $\v P$, spin $J=2, J_z=2$  is given   by
\be |{\cal T}_2, 2,\v P>={1\o2 \sqrt3}\int \prod_{i=1}^4d\v r_i\Psi_{\cal T}(\v r_1,\v r_2 ,\v r_3,\v r_4, \v P ) 
\tilde A_{+,a}^\dag(\v r_1) \tilde A_{+,b}^\dag(\v r_2) \tilde B_{+,a}^\dag(\v r_3) \tilde B_{+,b}^\dag(\v r_4)|0>\ ,\label{6}\ee 
while the state with $J=1$, $J_z=1$ and momentum $\v P$ is given by
\bea&&|{\cal T}_1, 1,\v P>={1\o2 \sqrt3}\int \prod_{i=1}^4d\v r_i \Psi_{\cal T}(\v r_1,\v r_2 ,\v r_3,\v r_4, \v P )\nn 
\[\tilde A_{+,a}^\dag(\v r_1)\tilde A_{+,b}^\dag(\v r_2)\tilde B_{+,a}^\dag(\v r_3) \tilde B_{-,b}^\dag(\v r_4)-\tilde A_{+,a}^\dag(\v r_1)\tilde A_{-,b}^\dag(\v r_2) \tilde B_{+,a}^\dag(\v r_3)\tilde B_{+,b}^\dag(\v r_4)\]|0>\ ,\label{6'}\eea 
and the state with $J=0$ and momentum $\v P$ is
\bea &&|{\cal T}_0,\v P>={1\o 6}\int \prod_{i=1}^4d\v r_i \Psi_{\cal T}(\v r_1,\v r_2 ,\v r_3,\v r_4, \v P )\nn
[\tilde A_{+,a}^\dag(\v r_1) \tilde A_{+,b}^\dag(\v r_2) \tilde B_{-,a}^\dag(\v r_3) \tilde B_{-,b}^\dag(\v r_4)+ \tilde A_{-,a}^\dag(\v r_1) \tilde A_{-,b}^\dag(\v r_2) \tilde B_{+,a}^\dag(\v r_3) \tilde B_{+,b}^\dag(\v r_4) \nn 
- \tilde A_{+,a}^\dag(\v r_1) \tilde A_{-,b}^\dag(\v r_2) \tilde B_{+,a}^\dag(\v r_3) \tilde B_{-,b}^\dag(\v r_4)-\tilde A_{-,a}^\dag(\v r_1) \tilde A_{+,b}^\dag(\v r_2) \tilde B_{+,a}^\dag(\v r_3) \tilde B_{-,b}^\dag(\v r_4) ]|0>\ .
\label{7}\eea

\section{The Drell-Yan production cross sections.}
\bigskip

 The parton-parton luminosity is given by the  convolution integral of two parton distribution functions which depend on the Bjorken variables $x_i\ ,\ i=1,2$  of the two partons and on the scale which is usually identified with the invariant mass of the two parton system, that is, in the collinear case, with $M^2\equiv s  x_1 x_2$ where $\sqrt s$ is the proton-proton collision energy. The convolution integral is in fact the integral over the gluon pair rapidity $y=\ln (x_1/x_2)/2$, and hence over the final  particle rapidity, of the product  of the parton distribution functions multiplied by the acceptance function of the experimental apparatus. The  product  of the parton distribution functions (PDF's)  is usually called the differential parton-parton luminosity and is a function of $y$ and $M$.  In our case the partons which dominate the production mechanism are gluons.
Thus we need the gluon-gluon luminosities for the LHC at $\sqrt s=13\ TeV$  which can be computed from the   literature,  in particular, from the references  \cite{8}, \cite{1l} and \cite{ball} \footnote{PDFs are released by individual groups as discrete grids of functions of the Bjorken-x and energy scale M. The LHAPDF project  \cite{0l} maintains a repository of PDFs from various groups in a new standardized LHAPDF6 format, additional  formats such as the CTEQ PDS grid format \cite{1l} are also in use.}. 

In the Appendices D.1 and D.2  we compute  in the ${\cal T}$ rest frame the transition amplitudes ${\cal T}\to 2 g$ and ${\cal T}\to q\bar q$  reminding that, if the parton masses are negligible, the invariant transition amplitude is given by 
$${\cal M}_{{\cal T}\to 2p}=-4\ (\pi \ M_{\cal T})^{3\o2} T_{{\cal T}\to 2p}\ .$$
Once these transition amplitudes are known we can compute the Drell-Yan production cross sections of ${\cal T}$ at the LHC in terms of the parton-parton luminosities.

Forgetting for simplicity the dependence on transverse momenta and denoting by $\ep_1$ and $\ep_2 $ the parton energies  we get for a single ${\cal T}$ component at its threshold 
$$\bar\s_0 =16 \pi^4 \ M_{\cal T} \overline{|T_{{\cal T}\to 2p}|^2} \d(4\ep_1\ep_2-M_{\cal T}^2) 
\ ,$$ where we have introduced the average over the parton helicities, momenta and colors. Lorentz invariance implies that the ${\cal T}$ production cross section does not depend on its rapidity.

If $f_g(x_i, \mu)$ is the $g$-partonic density the integrated ${\cal T}$ production cross section is \bea && \s_{{\cal T}} \simeq 16 \pi^4 \ M_{\cal T} \overline{|T_{{\cal T}\to 2p}|^2}\int_0^1 d x_1d x_2 \Theta (x_1, x_2)f_g(x_1, M_{\cal T})f_g(x_2, M_{\cal T}) \d(x_1x_2s-M_{\cal T}^2)\nn = 16 \pi^4 \ M_{\cal T} \overline{|T_{{\cal T}\to 2p}|^2}{1\o s}\int_{-\ln{\sqrt s\o M_{\cal T}}}^{\ln{\sqrt s\o M_{\cal T}}} d y\ \Theta (y)f_g({M_{\cal T} \o\sqrt s }e^y, M_{\cal T})f_g({M_{\cal T} \o\sqrt s }e^{-y}
, M_{\cal T})
\ ,\label{S1} \eea where the $\Theta$ functions account for the rapidity acceptance of the experimental apparatuses.

As a matter of fact in reference \cite{8} one finds the full luminosities corresponding for us to $|y|< 6.5=\ln{\sqrt s\o M_{{\cal T}_b}}$. On the contrary we must consider that at the CMS the luminosity is restricted to $|y|< 2$ while at the LHCb it is restricted to the interval $2\leq y\leq 4.5 $.

Disregarding the acceptance reductions due to cuts on the transverse momenta applied by the experimental apparatuses \cite{3}, for simplicity and for the incompleteness of our information, we can evaluate the gluonic luminosity which is given by
\be {\p L\o\p M^2}(M)={1\o s}\int_{-\ln{\sqrt s\o M}}^{\ln{\sqrt s\o M}} d y\Theta (y)f_g({M \o\sqrt s }e^y, M)f_g({M \o\sqrt s }e^{-y}
, M)\ .\label{S2}\ee
If the partons are light quarks the luminosity lessens by two orders of magnitude that of gluon pairs.

Using the mean square transition amplitudes given in Eq.s (\ref{tm0}) , (\ref{tm2}) and (\ref{tm2q}) in the Appendices D.1 and D.2 and using Eq. (\ref{S1}) together with the values of luminosities given in the third  and sixth rows of Table 1, we compute the production cross sections of the most interesting states 
of tetra-Quarks. Taking into account that ${\cal T}_2$ has the spin multiplicity equal to 5 we get the contribution of gluons to the cross section \be \s_{{\cal T}_2} \simeq 80\ \pi^4 \ M_{\cal T} \overline{|T_{{\cal T}_2 \to 2g}|^2} {\p L_{2p}\o\p M_{\cal T}^2}(M_{\cal T})\simeq 13\ \s_{{\cal T}_0} \ ,\label{sprod}\ee where, from Equation (\ref{tm2}), we have \be\overline{|T_{{\cal T}_2 \to 2g}|^2} \simeq 2.65\ 10^{-3} {\a_S^4\o \ m_Q^{10} \ d^6\d^{3 }}\ .\label{tinv}\ee
For the above-mentioned reasons concerning the light quark luminosity and considering Eq. (\ref{tm2q}) we can conclude that the light quark contributions to the cross section are negligible.

\subsection{The ratio ${\cal R}$ of the double $\mu\bar\mu$ pair  ${\cal T}$ resonant production to that from Quarkonium pairs.}

The possibility of detecting a ${\cal T}$ critically depends on the background in the chosen decay channel. According to \cite{4} the best choice is the channel of two $\mu\bar \mu$ pairs, each pair having a center of mass energy not far from the mass of the corresponding Quarkonium ${\cal Q}$. 
At least in the charm case   the background is mainly due to the production of pairs of ${\cal Q}$ both decaying into $\mu\bar\mu$ pairs. Therefore the signal to background ratio is strictly related to the ratio ${\cal R}$ of the rate of two $\mu\bar\mu$ pair production from the tetra-Quark decays to that from the  pairs of Quarkonia decaying into $\mu\bar\mu$ pairs.
If in the calculation of ${\cal R}$ we distinguish the contribution from tetra-Quarks of different spin, that is we write ${\cal R}={\cal R}_2+{\cal R}_0$, for each term we have
\be {\cal R}_J={\s_{{\cal T}_J}\o\s_{2{\cal Q}}}{B_{{\cal T}_J\to 2 \mu 2\bar\mu}\o B_{{\cal Q}\to 2\mu}^2}\ ,\label{rmu}\ee 
where $B_{{\cal T}_J\to 2 \mu 2\bar\mu}$ and $B_{{\cal Q}\to 2\mu}$ are the decay branching  ratios into $\mu\bar\mu$ pairs of the tetra-Quark and of the Quarkonium.

The LHCb Collaboration has measured, albeit with big uncertainties, the ratio ${\cal R}$ for the tetra-charm \cite{4}, and \cite {s2j} the production cross section of Quarkonium pairs $\s_{2 J/\psi}$. 

Our calculations of ${\cal R} $ for ${\cal T}_2$ 
and ${\cal T}_0$ are based on Eq. (\ref{2ymm}) in Appendix B.1 , that is 
$$\C_{{\cal T}_2\to{\cal Q} \mu \bar\mu}\simeq
{ 2^{19\o2} \ \ d^3\ \d^3 D^6 \o 3 (4d^2\d^2+D^2(d^2+2\d^2))^3 }\C_{{\cal Q}\to \mu\bar\mu}\simeq 4\ \C_{{\cal T}_0\to{\cal Q} \mu \bar\mu}\ ,$$ from which, assuming vector meson dominance, we have the following decay rate of ${\cal T}_2$ into two $\mu\bar\mu$ pairs 
$$\C_{{\cal T}_2\to2 \mu 2\bar\mu}\simeq
{ 2^{19\o2} \ \ d^3\ \d^3 D^6 \o 3 (4d^2\d^2+D^2(d^2+2\d^2))^3 }{\C_{{\cal Q}\to \mu\bar\mu}^2 \o \C_{\cal Q}}\ .$$ 
Therefore the ${\cal R}_J$ ratios are given by
\be {\cal R}_2\simeq { 2^{19\o2} \ \ d^3\ \d^3 D^6 \o 3 (4d^2\d^2+D^2(d^2+2\d^2))^3 } {\s_{{\cal T}_2}\o\C_{{\cal T}_2}}{\C_{\cal Q}\o \s_{2{\cal Q}}}\simeq 4.4\ {\cal R}_0\ .\label{rmu2}\ee
The Quarkonium widths $\C_{\cal Q}$ can be found in Particle Data Books while the cross sections $\s_{2{\cal Q}}$ are given,  at the LHC energy $ \sqrt s= 13 \ TeV$, in \cite{s2j} for the charmonium pairs and \cite{3} for the production of the bottomium pairs. 

In order to complete the calculation we need the tetra-Quark widths $\C_{\cal T}$ which we approximate to their strong decay rates at the lower orders in $\a_S$. The ${\cal T}$ decay rates ${\cal T}_J\to Q\bar Q 2g$ and ${\cal T}_2\to Q\bar Q q\bar q$ are computed in Appendix C, Eq.s (\ref{bb2g}), (\ref{bbqq}) , (\ref{bb2g0}) and (\ref{bbqq0}).
The third order corrections are expected to be $10 \%$ corrections.
Therefore, we shall limit our calculation to the second order in $\a_S$ and use Eq. (\ref{2tot})
$$
\C_{{\cal T }_2}\simeq {23\ \a_S^2\o m_Q^2\ (d^2+2\d^2)^{3\o2}} \simeq 3\ \C_{{\cal T }_0}\ .$$
Using all the above data and equations together with Eq.s (\ref{sprod}) and (\ref{tinv}) where we replace $M_{\cal T}$ with $4 m_Q$, we find
\be {\cal R}_2\simeq 1.2\ 10^3\ { (d^2+2\d^2)^{3\o2} \o(4d^3\d^2+d\ D^2(d^2+2\d^2))^3 } {\a_S^2(4m_Q^2)\ D^6\o m_Q^{7} }\ { \ \C_{\cal Q}\o \s_{2{\cal Q}}} {\p L_{2g}\o\p M_{\cal T}^2}\ .\label{rmu3}\ee

\bigskip

\subsection{The gluon-gluon luminosities.} 
\bigskip

\bigskip

The Drell-Yan production cross sections given in Equation (\ref{sprod}) and thence the $\eta_Q$ and ${\cal T}$ production cross sections and the signal to background ratios ${\cal R}$ given in Equation (\ref{rmu3}) are proportional to the gluon-gluon luminosities $\p L/\p M^2 (M)$ as shown in Equation (\ref{S2}). 

Thus we need the LHC gluon-gluon luminosity densities at $\sqrt s=13\ TeV$ which are given in the literature, in particular, in the references \cite{8}, \cite{1l} and \cite{ball}.

\begin{table}[!h]
\centering
\begin{tabular}{|l| c | c | c | c |}
\hline
Apparatus-coll. $\setminus$ Mass & $3\ GeV$ & $6\ GeV$ & $10\ GeV$ & $20\ GeV$\\
\hline
LHCb-MSTW.n & $4\ 10^{10} $ & $6.14\ 10^9 $ & $1.95\ 10^9 $ & $2.9\ 10^8 $ \\ 
\hline
LHCb-NNPDF.n & $5.74\ 10^{9} $ & $2.46\ 10^9 $ & $9.28\ 10^8 $ & $1.7\ 10^8 $ \\ 
\hline LHCb-CTEQ.n & $4.91\ 10^{9} $ & $2.43\ 10^9 $ & $9.58\ 10^8 $ & $1.88\ 10^8 $ \\ 
\hline
CMS-MSTW.n & $6.32\ 10^{10} $ & $1.24\ 10^{10} $ & $4.29\ 10^9 $ & $6.68 \ 10^8$ \\
\hline
CMS-NNPDF.n & $7.5\ 10^9 $ & $4.6\ 10^9 $ & $2 \ 10^9 $ & $4\ 10^8 $ \\ 
\hline
CMS-CTEQ.n & $8.1\ 10^{9} $ & $5.15\ 10^9 $ & $2.23\ 10^9 $ & $4.34\ 10^8 $ \\ 
\hline
\end{tabular}
\caption{Gluon-gluon luminosity table in picobarns as explained in the text below.}

\end{table}

In Table 1 we report in picobarn the luminosities seen by the apparatuses CMS and LHCb according to the data grills given by the Collaborations MSTW \cite{8}, NNPDF \cite{ball} and CTEQ \cite{1l} at the scales 3, 6, 10 and 20 GeV. Notice that at the scale 3 GeV the luminosities computed from the data given by the MSTW collaboration are about ten times larger than those computed by the Collaborations NNPDF and CTEQ, which essentially agree, while the MSTW data  decrease to less than the double of the same luminosities at the scale 20 GeV. Therefore, it is apparent that the extrapolations made by the MSTW Collaboration are based on different evolution equations from those used by the other two Collaborations.  In fact, minor  differences in the evolution equations may give appreciable differences in the luminosities at low scale because the extrapolations are based on the experimental results  at the masses of the vector boson Z and  of the Higgs particle. Notice also that, due to the different acceptances, the luminosities at the CMS are twice as much as those at the LHCb. 

A  criterion of choice among the given luminosities, in particular at low scale,  can be based on the comparison of  the published experimental value of   the  prompt production cross section of, e.g., the $\eta_c$ with the results of our calculations, given in Section 5, on the same process. This comparison gives    an indication of the gluon-gluon luminosity at the $3\ GeV $ scale.   

 The experimental value of $\s_{\eta_c}$ can be obtained from the two publications due to the LHCb Collaboration referred to in \cite{setac}.
The first of these  publications gives the prompt production cross section of a $J/\psi$ and  that of an $\eta_c$ with transverse momentum larger than $6.5\ GeV$. In the second, more recent,  publication it is given the ratio of the
$J/\psi$ to the $\eta_c$ prompt production cross sections. Considering both results one finds $\s_{\eta_c}= 25.4 \ \pm \ 7.7\ \mu b$.   
\footnote{These data give an indication of the transverse momentum distribution of the produced Quarkonia and hence of the two-gluon transverse momentum distribution at the mass of $3\ GeV$.  Assuming a Gaussian distribution we find a variance of $\overline{p_t^2}\sim 14\pm 2.5\ GeV^2$.}

Comparing the experimental results to our calculation we should find exact agreement if the gluon-gluon luminosity at the LHCb at $3 \ GeV$ were about $3\ 10^9\ pb$. This value  is lower  by a factor $1.6$ than the smaller value, that given by the CTEQ Collaboration, appearing in the first column of Table 1. Taking into account in particular the already discussed calculation uncertainties we can consider the result of this comparison a strong indication in favor of the data given by the CTEQ Collaboration that we shall use in all the cross section calculations.

 Unfortunately we have not been able to find among the LHC publications data about the prompt $\eta_b$ production cross section \footnote{In particular there is no reference to the $\eta_b$ in \cite{s2b} where the production cross section of bottom pairs is presented. We think that this is due to the low detection efficiency of gluon pairs at $10\ GeV$ due to the completely different hadronic conversion.}. Indeed, the knowledge of the prompt $\eta_b$ production cross section would have given us a further, more reliable, test of our calculations at the  $10 \ GeV$ scale.

Notice that an analogous comparison could be made for the ${\cal R}_c$ ratio at $6\ GeV$ getting the opposite result, however, taking into account the experimental together the calculation uncertainties, we prefer to postpone the discussion on this point to the next Section.

Therefore, in conclusion, we base our calculations on the gluon-gluon luminosities which are listed in Table 1 at the scales $M\sim 3\ GeV\ {\rm and } \ M\sim 6\ GeV$ at the LHCb and at the scales $M\sim 10\ GeV\ {\rm and } \ M\sim 20\ GeV$ at the CMS. These are based on the data given by the Collaboration CTEQ, \cite{1l} . 

On account of the approximations made in the construction of our non-relativistic scheme the difference between the data obtained from the references \cite{1l} and \cite{ball} turns out to be scarcely relevant.

\section{Numerical results on charm and bottom compound states}

\bigskip
In this Section we give our results and comment their meaning for what concerns the possible detection of tetra-bottom states.

We use the parameters given in the Sections 2 and 3, that is, for the charm flavor we have set
$m_c= 1.55\ GeV\ ,\ D_c= 1.6\ GeV^{-1}\ ,\ \d_c= 1.44\ GeV^{-1}\ ,\ d_c= 1.87\ GeV^{-1}$ and for the bottom $m_b= 4.73\ GeV\ , D_b= .8\ GeV^{-1}\ ,\ \d_b= .803\ GeV^{-1}\ ,\ d_b= 1.01\ GeV^{-1}$ together with the values of $\a_S (M^2)$ given in reference \cite{9}. 
In particular we have set in units $GeV^2$, $\a_S (2.4)=.32\ ,$ $\a_S (10)=.25\ ,$ $\a_S (22)=.21\ ,$ $\a_S (100)=.17\ .$ 

The first point we discuss concerns the reliability of the non-relativistic approximation for the scalar Quarkonia.
Using Equation (\ref{eta2f}) in Appendix B we find the decay widths into two gammas
$$ \C_{\eta_c\to 2 \c}= 7.2\ keV, \quad {\rm and} \quad \C_{\eta_b\to 2 \c}= .38\ keV\ .$$ Taking into account the full widths given by the Particle Data Book we find the $2\c$ branching ratios
$$ B_{\eta_c\to 2 \c}=4\ 10^{-4}\ ,$$ to be compared with the ratio $4.3\pm1.5\ 10^{-4}$ given by 
the Particle Data Book. As a matter of fact, the two values are perfectly compatible,  which is better than expected for the charm flavor.

In the bottom case, where the agreement between experiment and theory should be better, we find
$$ B_{\eta_b\to 2 \c}= 3.8 \pm 1.8\ 10^{-5}\ ,$$ but no experimental value is given because the 2$\c$ decay has not been seen.

Our calculations of the  $\eta_Q$ total decay width, that we identify with the two-gluon decay width, give the following results.

We begin from the $\eta_c$ whose width computed in Appendix C, Equation (\ref{eta2g}), is
\be \C_{\eta_c}={8 \a_s^2( m_Q^2)\o 3 \sqrt\pi m_Q^2\ D_{\cal Q}^3}\simeq25\ MeV\ ,\label{res1}\ee that we have compared in the previous Section with the datum given by the Particle Data Book (i. e. $17.3 \pm 2.5 \ MeV$) which is smaller but compatible with our result.

The same calculations for the $\eta_b$ give the following decay width 
\be\C_{\eta_b} \simeq 4.6\ MeV\ ,\label{res3}\ee which must be compared with the Particle Data Book value that is $10^{+5}_{-4}\ MeV$. Let us note that
both computed decay widths of the pseudoscalar Quarkonia reasonably agree with their experimental values.

Using furthermore the gluon-gluon luminosity values given in Table 1 we can compute the numerical values of the production cross sections of charm and bottom compound states. 

The $\eta_c$ production cross section $\s_{\eta_c}$ is obtained introducing into Equation (\ref{sprod}) the average square modulus of the transition amplitude to a gluon pair, which is 
\be\overline{|T_{\eta_c \to 2g}|^2} \simeq {\a_S^2( m_c^2)\o 192\ \ m_c^{4} \pi^{5\o2} D_c^3}\ ,\label{tinveta}\ee and is directly related to the above given value of $\C_{\eta_c}$.
Using  this equation we have
\be \s_{\eta_c}\simeq {\pi ^{3\o2}\ \a_S^2( m_c^2)\o 6\ m_c^3\ D_c^3}{\p L\o\p M^2}|_{LHCb}(2 m_c) \simeq 8\ 10^{-3}\ {\p L\o\p M}|_{LHCb}(3\ GeV)\simeq 39\ \mu b\ ,\label{res2}\ee while, as we have already said in the last Section, the cross section measured at the LHCb \cite{setac} is $\s_{\eta_c}\simeq 25.4\pm7.7\ \mu b$. Once again our value is compatible with the experimental one within  $2 \s$ despite the low value of the final particle mass and hence   the uncertainties in the gluon-gluon luminosity and those due to the relativistic corrections  and  to the transverse momenta corrections. Let us remind that, in particular for the charm data, our results are foreseen to give just the order of magnitude of the experimental values.

For the $\eta_b$ prompt production cross section   we get
\be \s_{\eta_b}\simeq {\pi ^{3\o2}\ \a_S^2( m_b^2)\o 6\ m_b^3\ D_b^3}{\p L\o\p M}|_{CMS}(2 m_b) \simeq 7.5\ 10^{-4}\ {\p L\o\p M^2}|_{CMS}(10\ GeV)\simeq 1.7 \ \ \mu b\ .\label{res4}\ee
We have not been able to find in the existing literature  any published experimental value of this cross section which is quite smaller than that of $\eta_c$.   . Apparently the reason for this difference is that, while the $\eta_c$  has been detected   through its decay into a proton-anti-proton pair, albeit this decay channel has  a branching ratio  of about $10^{-3}$, nothing is known about the decay of  the $\eta_b$ in the same channel.  Let us hope that some new result will soon appear.

Next we consider the tetra-Quark states beginning from the resonant production of two $\mu\bar\mu$ pairs from the ${\cal T}_2$ decay. We consider first the experimental result given by the LHCb Collaboration \cite{4} which refers to the signal to background ratio ${\cal R} $ for the tetra-charm, where the background is mainly due to the $\mu\bar\mu$ decay of a pair of vector Quarkonia. 

Our  results given  in  Equations (\ref{rmu2}) and  (\ref{rmu3}) show that the larger ratio is that associated with the $J=2$ state, therefore we limit our considerations to ${\cal R}_2 $. Using the data recalled at the beginning of the present Section and the $J/\psi$ width given by the Particle Data Books \footnote{The recent Particle Data Books give $\C_{J/\psi}\simeq 9.3\ 10^{-5} $ and $\C_{\Upsilon}\simeq 5.4\ 10^{-5} $.}, i.e. 
 $\C_{J/\psi}\simeq 9.3\ 10^{-5}\  ,$  we consider that the  value of double $J/\psi$ production cross section,  i.e. $\s_{2J/\psi}\simeq 1.5\pm .2 \ 10^{4}\ pb $, has been measured in the rapidity interval $2<y<4.5$  by the LHCb Collaboration and published  in \cite{s2j}.

From these data  and using  the gluon-gluon luminosity resulting from the data grills given by CTEQ Collaboration after the reduction due to the rapidity acceptance, we compute  \be {\cal R}_{c 2} \simeq 8 \ 10^{-13}{\p L\o\p M^2}|_{LHCb}(6\ GeV) \simeq ( 1.5\pm .16)\ 10^{-3}\ .\label{res5}\ee

 The LHCb Collaboration, selecting the  interval $6.2  < M_{di-J/\psi}   < 7.4 \ GeV$ for the invariant mass of the $J/\psi$ pairs, has shown the presence of possibly two tetra-charm resonances giving the ratio  ${\cal R}$ for the more apparent structure that we identify with that associated with $J=2$.
The values given in reference \cite{4}   vary between $ (1.1\pm .7)\ 10^{-2}$,  without any  selection of the total transverse momentum $p_T$, and $(2.6\pm 1.4)\ 10^{-2}$ for $p_T> 5.2\ GeV$. 

 In the same paper the  widths of  the most apparent  tetra-charm  resonance  is given, albeit with a wide uncertainty due to the possibility of interference of the resonance with the continuum  background.

Our result concerning ${\cal R}_2$, even being six times smaller than the experimental result, is compatible with  the experimental indications. Furthermore we should consider the calculation uncertainties of our result including those  due to the uncertain value of the gluon-gluon luminosity \footnote{Notice that using the luminosity given by the MSTW collaboration we should get ${\cal R}_2\simeq 4\ 10^{-3}$.}  and those related to the presence of   transverse momenta and to  the need of relativistic corrections.

In the bottom case we must refer to the results published  \cite{3} by the CMS Collaboration who have measured the production cross section of $\Upsilon$ pairs giving the value $\s_{2\Upsilon}\simeq 79\pm 20\ pb $. Then using  $\C_{\Upsilon}\simeq 5.4\ 10^{-5} $  we get
\be {\cal R}_{b 2}\simeq 2.65\ 10^{-11}{\p L\o\p M^2}|_{CMS}(20\ GeV)\simeq (1.1\pm .3) \ 10^{-2}\ ,\label{res6}\ee which is worth seven times   $  {\cal R}_{c 2}$. However in the bottom case ${\cal R}$ is much larger ( see \cite{3}) than the signal to background ratio for the tetra-Quark resonance because at the invariant mass of $20\ GeV$  the $\mu\bar\mu$ production rate is much  larger than that due to the double $\Upsilon\to \mu\bar\mu $ decay.
Still this result looks in agreement with what is shown in reference \cite{3}, Figure 7, which  has a significance of only one standard deviation. 

Perhaps an higher integrated luminosity might improve this agreement.

Thus, in order to verify the possibility of repeating for the bottom flavor at CMS the LHCb Collaboration measure for the charm, one has to consider beyond the ratios $ {\cal R}_{ 2}$ the value of the tetra-Quark production cross sections. From Equations (\ref{sprod}) and (\ref{tinv}) we get
\be \s_{{\cal T}_2} \simeq 108 \ {\a_S^4\o  \ m_Q^{9} \ d_Q^6\d_Q^{3 }} {\p L_{2g}\o\p M^2} (4 m_Q)\ \simeq 13\ \s_{{\cal T}_0} \ ,\label{sprod'}\ee 
thence, for the tetra-charm at the LHCb, we have
\be \s_{{\cal T}_{c 2}}\simeq 13\  \s_{{\cal T}_{c 0}} \simeq 6.42\ 10^{-5} {\p L_{2g}\o\p M^2}|_{LHCb}(6\ GeV)\simeq .16\ \mu b\ ,\label{res7}\ee while for the tetra-bottom at CMS we have
\be \s_{{\cal T}_{b 2}}\simeq 13\ \s_{{\cal T}_{b 0}} \simeq 1.5\ 10^{-7} {\p L_{2g}\o\p M^2}|_{CMS}(20\ GeV) \simeq 64 \ pb\ ,\label{res8}\ee which is about three  orders of magnitude smaller than the tetra-charm cross section.

It is also important to compute the value of the ${\cal T}$ width that, from 
Equations (\ref{2tot}) and (\ref{0tot}), turn out to be
\be
\C_{{\cal T }_{c 2}} \simeq 3\ \C_{{\cal T }_{c 0}} \simeq 27 \ MeV\ ,\label{res9}\ee while for the tetra-bottom we have 
\be \C_{{\cal T }_{b 2}} \simeq 3\ \C_{{\cal T }_{c 0}}\simeq 8 \ MeV\ .\label{res10}\ee
Remember that the width of a component (perhaps the $J=2$ one) of the tetra-charm has been measured by the LHCb Collaboration \cite{4} getting $\C_{{\cal T }_{c ?}}=80\pm 52 \ MeV$ assuming no interference with the background and $\C_{{\cal T }_{c ?}}=168\pm 102\ MeV$ assuming interference. Due to these big uncertainties our result is compatible with both experimental data.

Further results concerning the decay properties of the tetra-Quarks are obtained from Equation (\ref {0ymm}), getting for the tetra-charm
\bea&&\C_{{\cal T}_{c 2}\to\Upsilon \mu \bar\mu}\simeq 4\ \C_{{\cal T}_{c 0}\to\Upsilon \mu \bar\mu}\simeq .69\ \C_{J/\psi\to \mu\bar\mu}\simeq 3.7\ keV\ ,\nn
\C_{{\cal T}_{c 2}\to2 \mu 2\bar\mu}\simeq 4\ \C_{{\cal T}_{c 0}\to2 \mu 2\bar\mu}\simeq .22\ keV\ ,\label{res11}\eea 
while for the tetra-bottom we get
\bea&&\C_{{\cal T}_{b 2}\to\Upsilon \mu \bar\mu}\simeq 4\ \C_{{\cal T}_{b 0}\to\Upsilon \mu \bar\mu}\simeq .49\ \C_{\Upsilon\to \mu\bar\mu}\simeq .66\ keV\ ,\nn
\C_{{\cal T}_{b 2}\to2 \mu 2\bar\mu}\simeq 4\ \C_{{\cal T}_{b 0}\to2 \mu 2\bar\mu}\simeq 16\ eV\ ,\label{res12}\eea
from which it clearly appears that the tetra-bottom detection in the decay  into two $\mu\bar\mu$ pairs is much more difficult than the tetra-charm detection. Indeed, from Equations (\ref{res8}), (\ref{res10}) and (\ref{res12}), the resonant production cross section in two $\mu\bar\mu$ pairs due to the tetra-bottom turns out to be $\s_{{\cal T}_{b 2}\to2 \mu 2\bar\mu}\simeq .18 \ fb$. The CMS Collaboration trying to measure this cross section has given in reference \cite{3} an upper bound of $5\ pb$, more than four orders of magnitude above our result. However,  assuming our cross section value and considering that \cite{3} refers to an integrated luminosity of about $36\ fb^{-1}$, what appears in Figure 7 in   reference \cite{3} seems reasonable. Indeed,  taking into account the high $\mu\bar\mu$ pair detection efficiency   of the apparatus, our result
 would correspond to about $7$ events

We have already remarked that the tetra-Quark decay rate into a single $\mu\bar\mu $ pair together with a $Q\bar Q$ pair, which is computed in Equations (\ref {2bbmm}) and (\ref {0bbmm}), should be for the ${\cal T}_{c }$
\be \C_{{\cal T}_{c 2}\to Q,\bar Q, \mu \bar\mu}\simeq 4\ \C_{{\cal T}_{c 0}\to Q,\bar Q, \mu \bar\mu}\simeq 4\ keV\ ,
\label{res13}\ee 
and for the ${\cal T}_{b }$
\be \C_{{\cal T}_{b 2}\to Q,\bar Q, \mu \bar\mu}\simeq 4\ \C_{{\cal T}_{b 0}\to Q,\bar Q, \mu \bar\mu}\simeq .64\ keV \ .
\label{res14}\ee 
Thus, the resonant production cross sections in the considered channel turn out to be for ${\cal T}_{c }$ 
\be \s_{{\cal T}_{c 2}\to c,\bar c, \mu \bar\mu}\simeq 5.9\ \s_{{\cal T}_{c 0}\to c,\bar c, \mu \bar\mu}\simeq 23 \ pb\ ,
\label{res15}\ee and for
${\cal T}_{b }$ 
\be \s_{{\cal T}_{b 2}\to b,\bar b, \mu \bar\mu}\simeq .58 \  \s_{{\cal T}_{b 0}\to b,\bar b, \mu \bar\mu}\simeq 5.1 \ fb\ ,\label{res16}\ee which, with the present integrated luminosity, should correspond to about $200$ events.

However, the detection a tetra-bottom state in the channel $\mu\bar\mu \ b\bar b$  depends on the efficiency of the detection of a $b \bar b$ pair with an invariant mass of less than $1\ GeV$ above threshold \footnote{See footnote 2 and consider that the Quarkonium state $\Upsilon (4S)$, which mainly decays into a  pair of beauty particles,  has not been detected at the LHC.  } and with an average squared transverse momentum  about $40\  GeV^2$. The ${\cal T}$ detection efficiency in this channel would be favorite if  it were better than $2.7\ 10^{-2}$.  That is, the single beauty particle detection efficiency should be better than $.16 $. This looks very difficult to reach considering the very low statistics of the production process.

Therefore, we can conclude this analysis asserting that the most reliable channel where the tetra-bottom can be detected  seem to be that of two $\mu\bar\mu$ pairs.

\bigskip

\section{Conclusions.}

The aim of this paper is to present a coherent and possibly complete description of the phenomenology of the compound states of heavy quarks, in particular, of charm and bottom quarks. The paper is based on a non-relativistic Quantum Field Theoretical approach to the dynamics and on the evaluation of the gluon-gluon luminosities at the $13\ TeV$ LHC.

First of all, it must be remarked that the results should not be considered high precision results. This is due to four different factors. These are the relativistic corrections which are systematically disregarded, even for the charm flavor for which these corrections might be larger, and  the poor knowledge of the parton-parton LHC luminosities in the region of interest, i.e. between $3\ {\rm and }\ 20\ GeV$. Furthermore, we consider the production processes as purely collinear forgetting transverse momenta which are of the order of the particle masses, and we neglect higher order QCD perturbative corrections and hadronization corrections.

Some of these corrections should need very difficult calculations, some other are poorly known.
Therefore, the reliability of our results should be limited to their order of magnitude. However, as it clearly appears from the experimental results on the charm tetra-quarks and quarkonia the experimental data suffer from uncertainties ranging from ten to fifty percent. This is mainly due to the production processes being very rare and the detection efficiencies very low, in particular, at the LHC.

Despite these difficulties in the cases where a direct comparison of our results with the experimental data is possible, that is for the $\eta_c$ production cross section, width and $2\ \c$ decay rate, our results are in reasonable, in one case even surprising, agreement with the experimental results. There is also agreement on the decay width of the $\eta_b$. 

Concerning the recently uncovered tetra-charm at the LHCb a reasonable agreement exists for the data on the two $\mu\bar\mu$ pairs  decay width and the ratio of the tetra-charm to double $J/\psi$ production and decay rates. However, this agreement might be due to the poor quality of the present experimental data.

Therefore, despite the roughness of our approximations, we think, at the moment, useless to try to improve our calculations.

Our results on the tetra-bottom production and detection, which are the main purpose of the present paper, show the remarkable difficulties of the detection. As a matter of fact, the lack of the detection of the well-known $\eta_b$ at the LHC and the fact that the production cross section of $\Upsilon$ pairs at the CMS is about two hundred times smaller than that of $J/\psi$ pairs at the LHCb on a smaller rapidity interval gives a significant sign of the difficulty of the detection of any multi-bottom state. Our result for the ratio of the production cross section of a single tetra-bottom state $ \s_{{\cal T}_{b }}$ at the CMS to that of the tetra-charm state $ \s_{{\cal T}_{c}}$ at the LHCb, whose amount  is about $4 \ 10^{-4}$, confirms the mentioned difficulty.

Thus, what we can do is try to single out the better detection channel. Considering that the hard version of the most important decay channel is $b\bar b\ 2g$, that is, a pair of gluons together a $b\bar b $ pair and that, after hadronization, the two gluon jets with total invariant mass near $10\ GeV$ correspond to the $\eta_b$ decay products  while  a pair $b\bar b$ near threshold \footnote{And momentum difference of less than one $GeV$.} should hadronize as   the decay products of a $\Upsilon (4S)$, we think that one should have recourse to another decay channel due to the low detection efficiency of this channel.

For this reason, we have considered the case where the gluon pair is replaced by a $\mu\bar\mu$ where the muon transverse momenta might reach six-seven GeV while that of the bottoms might be roughly the same of that in the $b\bar b\ 2g$ channel. 
From our calculations the branching ratio of the decay into this channel is $B_{{\cal T}\to b\bar b\mu\bar\mu}\sim 10^{-4}$, thus this channel is convenient if the selection efficiency of the pair of gluon jets discussed above is worse than $10^{-4}$. Otherwise, the main channel ($b\bar b\ 2g$) should remain the more convenient.

The last possibility we have considered is the decay into two $\mu\bar\mu$ pairs, the channel in which the tetra-charm has been detected at the LHCb \cite{4} and the tetra-bottom has been looked for at the CMS \cite{3} . In this channel the computed decay branching ratio is worth $B_{{\cal T}\to 2\mu 2\bar\mu}\sim 2\ 10^{-5}$ and the resonant production cross section we have computed is about $\s_{{\cal T}_{b 2}\to2 \mu 2\bar\mu}\simeq .18 \ fb$. As mentioned in the previous Section this seems to be the favored channel for the tetra-bottom discovery due to the low  detection efficiency of the  channels with higher decay branching ratio, but, before a reasonable detection be possible, LHC has to reach an integrated luminosity of at least $500\ fb^{-1}\ .$

As a final comment we note that  we have only discussed the results concerning ${\cal T}_2$ and ${\cal T}_0$ for which we have quoted the ratios of our results to the ${\cal T}_2$ ones. In particular we have forgotten the ${\cal T}_1$ state because, in our approximation, its production cross section vanishes at the LHC.

\bigskip

\bigskip

\centerline{\Large\bf   Appendices}

\appendix
\section{Interactions in the non-relativistic limit.}

Beyond the evaluation of the production cross section of tetra-Quarks our work has been mainly devoted to the study of the properties of the Quarkomium and tetra-Quark states. We have used the same description of the tetra-Quark states already used in \cite{5} and \cite{6} assuming a non-relativistic structure based on the wave functions given in Eq.s (\ref{1}) and (\ref{2}) and on the fact that the Quarkonium radii $D_c$ and $D_b$ are sufficiently larger respectively than $m_c^{-1}$ and $m_b^{-1}$. 

We compute the transition amplitudes involved into the calculations of the Drell-Yan production cross sections and of the main hard decay processes using for each process the corresponding effective Hamiltonian which is identified in analogy with that of beta decay by the Feynman diagrams associated with the relevant interactions.

The effective Hamiltonians are related to the relevant tree approximation diagrams which should be dominant due to the momentum tranfers of the order of magnitude of the heavy Quark masses. Furthermore, they are computed in the rest frame of the initial heavy Quark compound pushing the non-relativistic approximation to the limit where the initial heavy Quarks are considered at rest while the light decay products, either gluons of light quarks or leptons, have light-like momenta.

Presenting the details of our calculations we must remind the structure of the non-relativistic spinor field operator.

Choosing the $\c$ matrices in the standard representation we have for the non-relativistic spinor field
\bea &&\Psi(\v r)=[\tilde A_\pm(\v r)-i\c_2\tilde B_\pm^\dag(\v r)]U_\pm=\int {d\v p\o(2\pi)^{3\o2}}
e^{i\v p\.\v r}[ A_\pm(\v p)-i\c_2 B_\pm^\dag(-\v p)]U_\pm
\ ,\nn \bar\Psi(\v r)=\bar U_\pm[\tilde A^\dag_\pm(\v r)+i\c_2 \tilde B_\pm(\v r)] =\bar U_\pm\int {d\v p\o(2\pi)^{3\o2}}
e^{i\v p\.\v r}[ A^\dag_\pm(\v p)+i\c_2 B_\pm (-\v p)]\ ,\label{campo}\eea where
$$U_\pm = \(\matrix{w_\pm\cr0\cr}\) \quad\ ,\quad V_\pm =\(\matrix{0\cr w_\pm\cr}\)\ .$$ and $$w_+=\( \matrix {1 \cr 0\cr}\)   \quad\ ,\quad w_-=\(\matrix{0\cr 1\cr}\)  \ .$$
The purpose of these Appendices consists in characterizing the effective Hamiltonians appropriate for the considered processes.

Thus, for example, to the first order in $\a$, the effective Hamiltonian accounting for the annihilation of a pair of heavy spinors (for example the Quarks $b\bar b$) at rest into a pair $\mu\bar\mu$ is given, neglecting the contribution of the exchange of a $Z$ intermediate boson and summing over the Quark color,  by 
\be H_I={4\pi\a\zeta\o 3s} \int d\v r\(\bar\Psi_a(\v r)\c_\nu \Psi_a(\v r)\)(\bar \psi(\v r) \c^\nu\psi(\v r))\sim {4\pi\a\zeta\o12m_Q^2}i \int d\v r [\tilde B_{s,a}(\v r)\tilde A_{s',a}(\v r) j_{ss'}(\v r)+h.c.]\ ,\label{elm}
\ee where $s$ is the squared annihilation tetra-momentum, $h.c.$ stays for the terms containing heavy Quark creation operators,
$\zeta= 1$ for the bottom and $\zeta=-2$ for the charm. We have set $$j_\mu(\v r)\equiv( \bar \psi (\v r) \c_\mu\psi (\v r) )\ ,\quad{\rm and \ hence}\ ,\quad \v j (\v r)\equiv( \bar \psi (\v r)\v \c\psi (\v r) )\ ,$$ where the field $\psi$ is  a relativistic spinor field, for example that of the $\mu$ particle. In QCD where the light spinors are quarks and hence carry color, we shall introduce and sum over the color indices as in$$j_\mu^\a(\v r)\equiv t^\a_{ab}( \bar \psi_a (\v r) \c_\mu\psi_b (\v r) )\ .$$

The Quark interaction Hamiltonian which has been used applying the variational method shown in the Introduction is
	\bea&& H_{int}= \a_S \int {d\v r d\v r'\o4|\v r-\v r'|}(\d_{da}\d_{bc}-{1\o3}\d_{dc}\d_{ba})(A^\dag_{a\s}(\v r)A_{b\s}(\v r)-B^\dag_{b\s}(\v r)B_{a\s}(\v r))\nn
(A^\dag_{c\s'}(\v r')A _{d\s'}(\v r')-B^\dag_{d\s'}(\v r')B_{c\s'}(\v r'))\ .\nonumber\eea

In the case of the first order in $\a$ annihilation of a pair $Q\bar Q$ with momenta $|\v p|\ll m_b$ only the destruction Quark and anti-Quark operators, e.g. $\tilde A_{+,c}(\v r)=\int d\v p\ A_{+,c}(\v p)\exp(i\v p\.\v r)/(2\pi)^{3/2}$, appear. Since the fields appearing in $\v j (\v r)$ carry much larger momenta than (anti-)Quarks, it is possible to replace $\tilde A_{+,c}(\v r)$ with $\tilde A_{+,c}(\v 0)$. Obviously the same holds true for the anti-Quark operator $\tilde B_{+,c}(\v r)$. As a consequence, the effective Hamiltonian may be written as
\be H_I\simeq {\pi\a\zeta\o3\ m_Q^2}i [\tilde B_{s,a}(\v 0)\tilde A_{s',a}(\v 0)\int d\v r \ j_{ss'}(\v r)+h.c.]\ .
\label{3} \ee 
Here, in Equation (\ref{elm}) and  in the rest of this paper we shall only be interested in the part of the effective Hamiltonian which induces destruction of the heavy Quarks. Since we shall often deal with rather cumbersome formulae any, however limited, simplification will be appropriate. Therefore, in the following, we shall
systematically omit any mention of the terms containing heavy Quark creation operators.

A further simplification follows from the fact that tetra-Quark states are color singlet states and hence the dependence of the final result on the color indices of the (anti-)Quark destruction operators  lies in an universal factor which can easily factorized out of the calculation. 
 This factor appears in the relation
\bea&&<0|\tilde B_{s_2 c_2} (\v r_2)\tilde A_{s_1 c_1}(\v r_1 ) \tilde B_{s_4 c_4} (\v r_4)\tilde A_{s_3 c_3}(\v r_3) 
|{\cal T}_J, J_z,\v P>=(\d_{c_1c_4}\d_{c_2c_3}-\d_{c_1c_2}\d_{c_3c_4})\nn<0|\tilde B_{s_2 } (\v r_2)\tilde A_{s_1 }(\v r_1 ) \tilde B_{s_4 } (\v r_4)\tilde A_{s_3 }(\v r_3) 
|{\cal T}_J, J_z,\v P>\ 
=( \d_{c_1c_4}\d_{c_2c_3}-\d_{c_1c_2}\d_{c_3c_4})\Psi_{\cal T}(\v r_1,\v r_2 ,\v r_3,\v r_4, \v P )\nn\Phi_{J,J_z}(s_i )\ ,
\label{defi}\eea in the second equation we have disregarded the  color index of the destruction operators which therefore are considered colorless.

Computing the matrix elements of the effective Hamiltonian for the decay of a tetra-Quarks, these matrix elements shall be divided in two factors using the factorization of the Fock space in a non-relativistic part containing the heavy Quark states and in a relativistic part spanned by the light particle states.

The first step of the calculation will consist in computing the operator valued matrix element $<H>_{NR}$ of the effective Hamiltonian between the initial state of the tetra-Quark and the heavy Quark component of the final state. The resulting $<H>_{NR}$ will be an operator in the Fock space of the light particles. The second step will consist in the calculation of the c-number matrix element of $<H>_{NR}$ between the vacuum state and the final state of the light particles.

\section{The electromagnetic decays}

The simplest test of our scheme consists in the  comparison of the decay rates of the $\eta_Q$'s into two gammas. This decay is induced by the effective Hamiltonian whose bottom  destruction component is
$$H_{2\c}=-i{\pi\ \a\ \zeta^2\o9\ m_Q}\tilde B_{s a}(\v 0) \tilde A_{s' a}(\v 0)\sum_\la[\la(\s_2)_{ss'}+(\s_2\v\s\.\v v)_{ss'}]\int {d\v q\o  q}\  a^\dag_\la (\v q)  a^\dag_\la (-\v q)\ .$$
The corresponding  transition matrix element is
$$<0| a _{\la_1} (\v q_1)  a _{\la_2} (\v q_2)\ H_{2\c}|\eta_Q, \v 0> =-i{\a\ \zeta^2\o  3^{3\o2} \ \pi^{5\o4}m_Q \ D_Q^{3\o2}} \d_{\la_1\la_2}\la_1{\d(\v k_1+\v k_2)\o k_1}\ ,$$ and the decay rate into two gammas is given by
\be \C_{\eta_Q\to 2 \c}={4\ \a^2\ \zeta^4\o 3^3\ \sqrt\pi \ m_Q^2\ D_Q^3}\ .\label{eta2f}\ee

\subsection{The decay ${\cal T} \to \mu\bar\mu{\cal Q}\ .$}

Here we compute the decay rate of the state ${\cal T}_2$ into two $\mu\bar\mu$ pairs assuming a two step process \cite{3}, in the first step 
${\cal T}_2$ decays electromagnetically into a $\mu\bar\mu$ pair and a Quarkonium state ${\cal Q}$, in the second step ${\cal Q}$ decays into another $\mu\bar\mu$ pair. 

Let us thus consider the matrix element of the effective Hamiltonian given in Eq. (\ref{3}) between the initial state of ${\cal T}_2$ at rest and a final Quarkonium ${\cal Q}$ state. The corresponding Fock vector states are given by $|{\cal T}_2 ,2, \v 0 >$ shown in Eq. (\ref{6}) and the Quarkonium state $|{\cal Q} ,1, \v P >$ given in Eq. (\ref{5}). 

In this first example we show the computation strategy that we shall apply many times in the following and is anticipated in Appendix A.

Considering the first of the two steps of the already discussed procedure we shall use the effective Hamiltonian given in Eq. (\ref{3}) together with Eq. (\ref{defi}) with $$ j_+(\v r)=(\bar\psi_\mu(\v r)(\c_x+i\c_y) \psi_\mu(\v r))\ , $$ obtaining the operator in the muon Fock space
\bea &&<{\cal Q},1, \v P |
H_I|{\cal T}_2,2,\v 0>_{NR}\sim
{-i \a\zeta\ 2^{5\o2}D^{3\o2}\ d^3\ \d^{3\o2} \o 3\ \pi^2 (4\d^2 d^2+D^2( d^2+2\d^2) )^{3\o2} m_b^2}e^{-{P^2 d^2\o4}}
\int d\v r\ j_+(\v r) e^{-i \v P\.\v r}\nn 
\ . \label{2y}\eea 
Then  computing the transition matrix element in the muon Fock space we get
\bea&&<{\cal Q},1, \v P, \v k_1, \lambda_1, \v k_2, \lambda_2|H_I|{\cal T}_2,2,\v 0 >={-i \a\ 2^{3\o2}D^{3\o2}\ d^3\ \d^{3\o2} \o 3\ \pi^2 (4\d^2 d^2+D^2( d^2+2\d^2) )^{3\o2} m_Q^2\sqrt{k_1k_2}}e^{-{P^2 d^2\o4}} \nn
( \bar u_{ \lambda_1}(\v k_1) \c_+ v_{ \lambda_2}(\v k_2))
\d(\v k_1+\v k_2)\ . \nonumber\eea
Therefore the decay rate of the ${\cal T}_2$ at rest into ${\cal Q} \mu\bar\mu$ is given by
\bea\C_{{\cal T}_2\to {\cal Q} \mu \bar\mu}\sim&&{ 2^{23\o2} \ \a ^2\zeta^2\ d^3\ \d^3 D^3 \o 27\ \sqrt\pi (4d^2\d^2+D^2(d^2+2\d^2))^3 m_Q^2}\nn\sim
{ 2^{19\o2} \ \ d^3\ \d^3 D^6 \o 3 (4d^2\d^2+D^2(d^2+2\d^2))^3 }\C_{{\cal Q}\to \mu\bar\mu}
\label {2ymm}\eea
Note that the final result  depends on the Quark charge only through $\C_{{\cal Q}\to \mu\bar\mu}$.

The decay rate of the ${\cal T}_0$ is easily computed following the same procedure and introducing a factor $3$ taking into account the spin degeneracy of the final ${\cal Q}$. Therefore we have
\be \C_{{\cal T}_0\to {\cal Q} \mu \bar\mu} \sim
{ 2^{15\o2} \ \ d^3\ \d^3 D^6 \o 3 (4d^2\d^2+D^2(d^2+2\d^2))^3 }\C_{{\cal Q}\to \mu\bar\mu} \simeq{1\o4} \C_{{\cal T}_2\to {\cal Q} \mu \bar\mu}
\label {0ymm}\ee 

The transition rate ${\cal T}_1\to{\cal Q} \mu \bar\mu$ vanishes due to charge conjugation invariance.

Now the decay rate into two $\mu\bar\mu$ pairs is easily computed multiplying the above results by the ${\cal Q}$ branching fraction into a $\mu \bar\mu$ pair. We get \bea&&\C_{{\cal T}_2\to 2 \mu 2\bar\mu}=4\C_{{\cal T}_0\to 2 \mu 2\bar\mu}={ 2^{19\o2} \ \ d^3\ \d^3 D^6 \o 3 (4d^2\d^2 +D^2(d^2+2\d^2))^3 }{\C^2_{{\cal Q}\to \mu\bar\mu}\o \C_{\cal Q}}\nn\simeq 240\ { \ d^3\ \d^3 D^6 \o (4d^2\d^2 +D^2(d^2+2\d^2))^3 }{\C^2_{{\cal Q}\to \mu\bar\mu}\o \C_{\cal Q}}\ . \label{24m}\eea

\subsection{ The decay ${\cal T} \to \mu\bar\mu Q \bar Q\ $ .} 

Then we consider the electromagnetic tetra-Quark decay into an open $Q\bar Q$ pair and a $\mu\bar\mu$ pair. We have suggested in the Conclusions Section that the decay into this channel might be interesting for the tetra-Quark detection.

The calculation of the decay rate ${\cal T} \to \mu\bar\mu Q \bar Q\ $ is analogous to that in the final channel ${\cal Q}\mu\bar\mu\ .$   One must replace the Quarkonium wave function by the product of two plane waves.

Thus beginning with ${\cal T}_2$, let us consider the production of a $\mu\bar\mu$ and $Q\bar Q$ pairs. 
In analogy with Eq. (\ref{2y}) we get
 the decay rate 
\be \C_{{\cal T}_2\to Q,\bar Q, \mu \bar\mu}\simeq
{ \a^2\zeta^2 \ 2^{11\o2}\o 3^3\ \sqrt\pi \ m_b^2(d^2+2\d^2)^{3\o2}}\simeq .95{ \a^2 \ \zeta^2 \o m_Q^2(d^2+2\d^2)^{3\o2}}\ ,
\label{2bbmm}\ee 
and  it is easy to see that
\be \C_{{\cal T}_0\to Q,\bar Q, \mu \bar\mu} ={1\o4}\C_{{\cal T}_2\to Q,\bar Q, \mu \bar\mu}\ .\label{0bbmm}\ee

  \section{The second order in $\a_S$.}
 
Now we consider the effective Hamiltonian for the second order in $\a_S$ annihilations of a $Q\bar Q$ pair in non-relativistic movement in, either a gluon pair, or a light $q\bar q$ pair. These correspond to the diagrams 
\footnote{Computing the effective Hamiltonian corresponding to a Feynman diagram we have to replace the external lines by the corresponding field operators. This changes the nature of the diagrams because the effective Hamiltonian matrix element corresponding to single diagram contains as many different terms as there are permutations of identical particles both in the initial and the final state. For example, in the scalar theory the first order invariant transition amplitude of two particles in two particles is just the coupling constant $\la$. 
On the contrary the effective Hamiltonian is just the first order interaction operator $\int d\v r\la \f^4(\v r)/4!\ .$ } shown in Figure 1.

\bigskip 

\begin{figure}[h]
\centering
\subfigure[]{\includegraphics[width=0.12\textwidth]{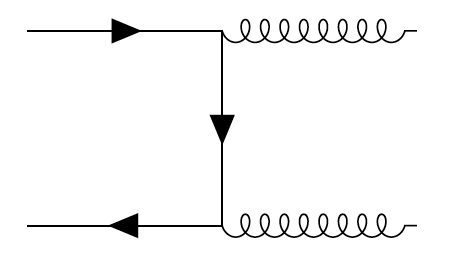}} 
\subfigure[]{\includegraphics[width=0.12\textwidth]{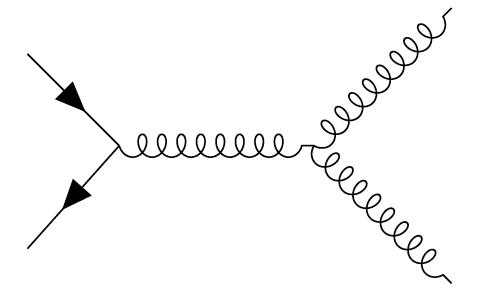}} 
\subfigure[]{\includegraphics[width=0.12\textwidth]{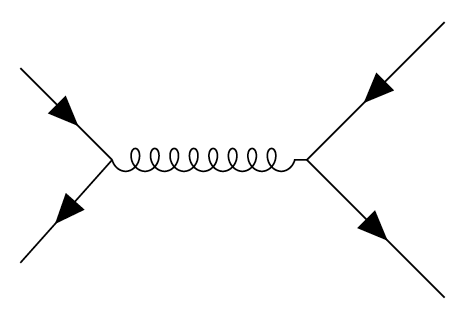}}
\caption{The order $\a_S^2$}
\label{fig:Order2}
\end{figure}
In the reference frame in which the initial heavy Quarks are in non-relativistic movement we 
 identify the heavy Quark  four-vectors entering into the diagrams from the left by
$$ p\sim p'\sim m_Q\( \matrix{\v 0 \cr  1 \cr }\)\ , {\rm and\ for\ outgoing\  light\  particles}\ , q\sim m_Q\( \matrix{\v v \cr  1 \cr }\) \ ,\quad q'\sim m_Q\(\matrix {-\v v \cr  1 \cr }\)\ .$$
Here and in the following we shall choose the gluon polarization vectors  setting, if $\v q$ is parallel to the $z$ axis, \be\v \ep_\la( \v q) = (\v x+i\lambda \v y)/\sqrt2\ ,\label{pol}\ee \ \footnote{Where $\v x$ and $\v y$ are also unimodular and parallel to the corresponding axes.} and using the standard convention  $\v \ep_\la(-\v q)=\v \ep^*_\la(\v q)$. 

It follows that, independently of the direction of the vector $\v v$ one has  
\be i \v v \.(\v \ep^*_{\la' }(\v q' )\wedge \v\ep^*_{\la }(\v q ))= \la \d_{\la\la'}\quad\ ,\quad \v \ep^*_{\la' }(\v q' )\. \v\ep^*_{\la }(\v q )=\d_{\la\la'}\ .\label{epsi}\ee

The decay amplitude of $\eta_Q $ into two gluons corresponds to diagram (a) which leads to 
the decay rate
\be \C_{\eta_Q\to 2g}={8 \a_s^2( m_Q^2)\o 3 \sqrt\pi m_Q^2\ D_{\cal Q}^3}\ .\label {eta2g}\ee

The third order contributions should not give substantial corrections.

It is also easy to compute the rate of the decay ${\cal T}_2\to Q\bar Q 2g$ which
 turns out to be
\be \C_{{\cal T }_2\to Q\bar Q 2g}={48\ \a_S^2\o\sqrt{2\pi}\ m_Q^2\ (d^2+2\d^2)^{3\o2}}\simeq 19{\a_S^2\o \ m_Q^2\ (d^2+2\d^2)^{3\o2}}\ .\label{bb2g}\ee 
Considering  the decay ${\cal T}_2\to Q\bar Q q\bar q$ 
we get the decay rate
\be \C_{{\cal T }_2\to Q\bar Q q\bar q}={2^{5}\ \a_S^2\o9\ \sqrt{2\pi}\ m_b^2\ (d^2+2\d^2)^{3\o2}}\ .\label{bbqq}\ee

The sum of the two rates gives  \footnote{We consider 3 light flavors.}
\be \C_{{\cal T }_2}\simeq {23\ \a_S^2\o m_Q^2\ (d^2+2\d^2)^{3\o2}}\ ,\label{2tot}\ee

If instead we consider the rate of the decay ${\cal T}_0\to Q\bar Q 2g$ 
we find the  rate
\be \C_{{\cal T }_0\to Q\bar Q 2g}={64\ \a_S^2\o3\sqrt{2\pi}\ m_Q^2\ (d^2+2\d^2)^{3\o2}}\simeq 8.6 {\a_S^2\o \ m_Q^2\ (d^2+2\d^2)^{3\o2}}\ .\label{bb2g0}\ee

On the contrary the rate of the decay ${\cal T}_0\to Q\bar Q q\bar q$ 
is
\be \C_{{\cal T }_0\to Q\bar Q q\bar q}={8\ \a_S^2\o3\ \sqrt{2\pi}\ m_Q^2\ (d^2+2\d^2)^{3\o2}}\simeq  {\a_S^2\o \ m_Q^2\ (d^2+2\d^2)^{3\o2}}\ .\label{bbqq0}\ee
Summing the two contributions to the second order decay rate we get
\be \C_{{\cal T }_0}\simeq 8.6{\a_S^2\o \ m_Q^2\ (d^2+2\d^2)^{3\o2}}\ ,\label{0tot}\ee which is roughly half of the second order decay rate of ${\cal T}_2$.

\section{Tetra-Quark transition to   a pair of light particles at the order $\a_S^4$.}

\begin{figure}[h]
\centering
\subfigure[]{\includegraphics[width=0.11\textwidth]{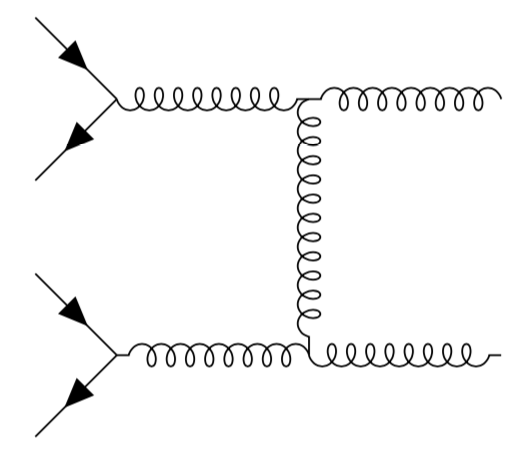}} 
\subfigure[]{\includegraphics[width=0.11\textwidth]{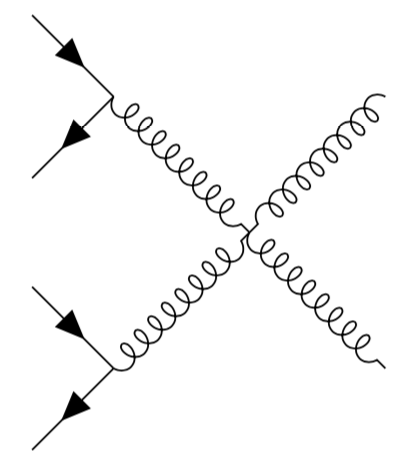}} 
\subfigure[]{\includegraphics[width=0.11\textwidth]{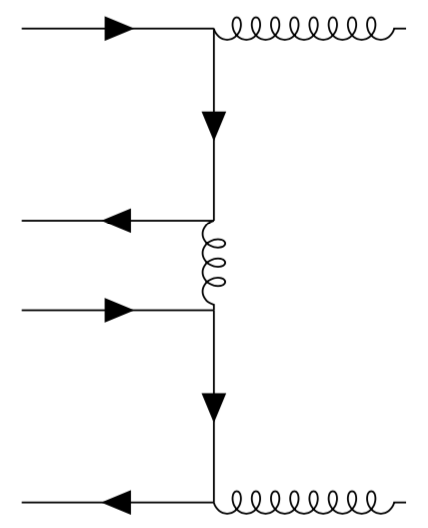}}
\subfigure[]{\includegraphics[width=0.11\textwidth]{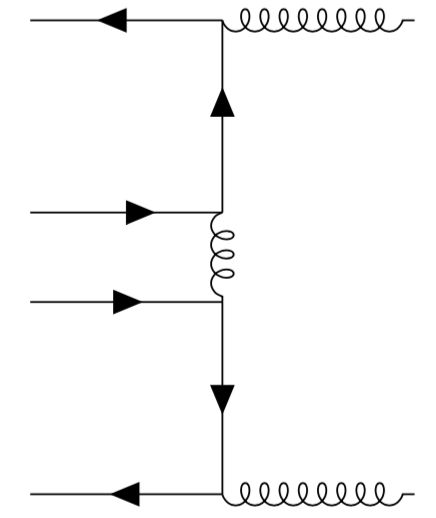}}
\subfigure[]{\includegraphics[width=0.11\textwidth]{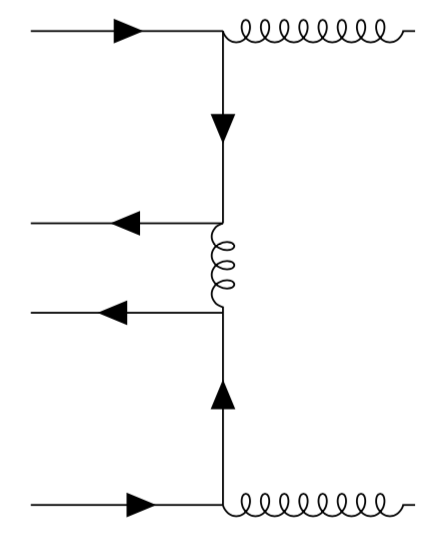}}
\subfigure[]{\includegraphics[width=0.11\textwidth]{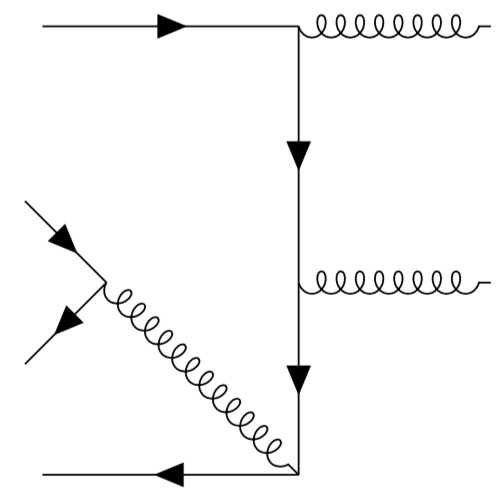}}
\subfigure[]{\includegraphics[width=0.11\textwidth]{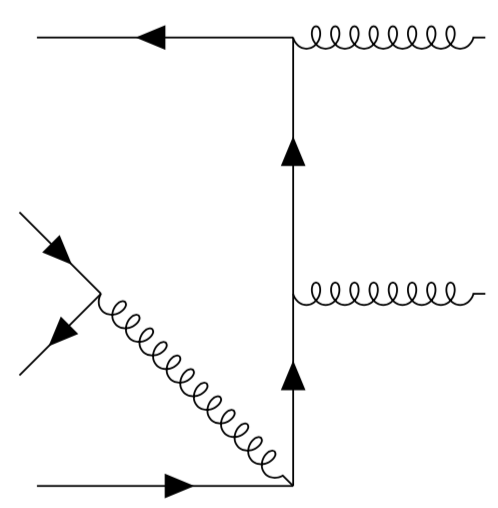}}
\subfigure[]{\includegraphics[width=0.11\textwidth]{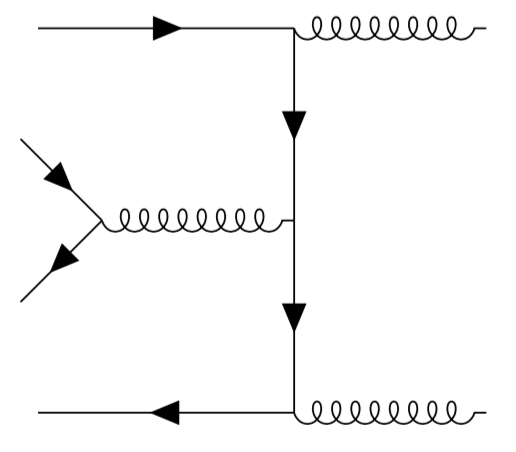}}
\end{figure}

\begin{figure}[h]
\centering
\subfigure[]{\includegraphics[width=0.11\textwidth]{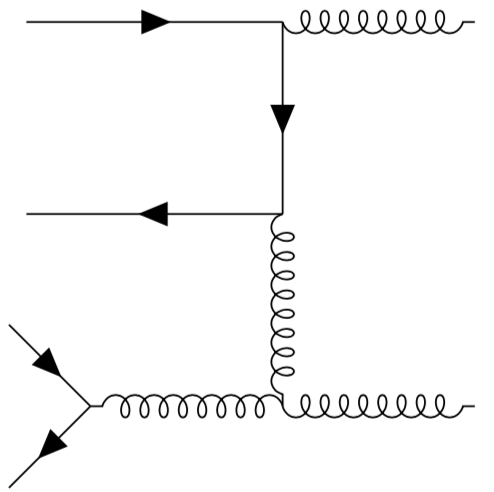}}
\subfigure[]{\includegraphics[width=0.11\textwidth]{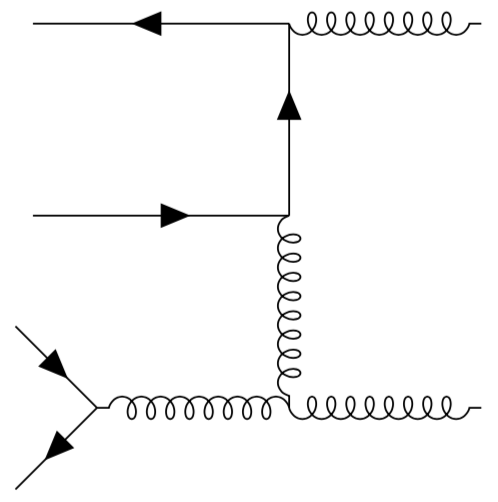}}
\caption{The transition ${\cal T}\leftrightarrow 2g$.}
\label{fig:Order4}
\end{figure}

Postponing to a further paper the study of the third order QCD transitions, now we discuss the transition amplitudes of a Tetra-Quark into a pair of light particles. 
This choice is justified in Section 5 where we have discussed the Drell-Yan mechanism.

 We have seen that the ${\cal T}$ production is due to a two parton fusion and that the case of two gluons is favored with respect to that of a light $q\bar q$ pair. Both are  fourth order processes.

We begin with the transition ${\cal T}\to 2 g$ which corresponds to the ten diagrams shown in Figure 2 where each diagram corresponds to a
 single terms of the effective Hamiltonian.

The momentum flow in all the diagrams is easily identified associating with the four heavy Quarks entering from the left the same momentum $$ p \sim m_Q\( \matrix {\v 0  \cr  1 \cr }\)\ . $$ On the contrary the outgoing gluons on the right have momenta
$$ q\sim 2 m_Q\(\matrix {\v v \cr  1 \cr }\)\ ,\quad q'\sim 2 m_Q\(\matrix {-\v v \cr  1 \cr }\)\ .$$

We must consider that the effective Hamiltonian acts on a color singlet state thus according to Equation (\ref{defi})  the dependence on the colors carried by the Quark fields factorizes in $\d_{c_1c_4}\d_{c_2c_3}-\d_{c_1c_2}\d_{c_3c_4}$ which must be contracted with the color factors of the diagram. The final result selects, as it should, the color singlet combination of the final gluons. The gluon field dependent part of the effective Hamiltonian consists in the two gluon creation operator  $$\int {d\v q\o 2q}a^\dag_{\la \a}(\v q) a^\dag_{\la' \b}(-\v q)\equiv \Xi_{\a\b\la\la'}\ ,{\rm and}\quad \d_{\a\b} \Xi_{\a\b\la\la'} \equiv \Xi_{\la\la'}\ .$$

We show the calculation of the effective Hamiltonian term corresponding to {\bf diagram a}  which we present in some details in order to give an explicit example of the simplifications discussed above.
\bea&&  H_a\sim {\ g^4\ \Xi_{\a\b\la\la'}\o(2m_Q)^6}f^{\a\c\eta} f^{\b\d\eta}t^\c_{c_2, c_1}t^\d_{c_4, c_3}\ \ep^*_{\la\eta }(\v q)\ep^*_{\la'\tau }(-\v q)(\bar\Psi_{2,c_2}(\v 0)\c_\mu \Psi_{1,c_1}(\v 0)) (\bar\Psi_{4,c_4}(\v 0)\c_\nu \Psi_{3,c_3}(\v 0))\nn
(g^{\eta\xi}2(q-p)^\mu-g^{\eta\mu}( 2 p+q)^\xi)
(-g^{\nu\tau}( 2p+q')_\xi  +g^{\tau\xi}2(p+q')^\nu)  
\nn 
\sim {\ g^4\ \Xi_{\a\b\la\la'}\o(2m_Q)^6}f^{\a\c\eta} f^{\b\d\eta}t^\c_{c_2, c_1}t^\d_{c_4, c_3}(\d_{c_1c_4}\d_{c_2c_3}-\d_{c_1c_2}\d_{c_3c_4})  \ep^*_{\la\eta }(\v q)\ep^*_{\la'\tau }(-\v q)(\bar\Psi_{2 }(\v 0)\c_\mu \Psi_{1 }(\v 0))\nn (\bar\Psi_{4 }(\v 0)\c_\nu \Psi_3(\v 0))
(g^{\eta\xi}2(q-p)^\mu-g^{\eta\mu}( 2 p+q)^\xi)
(-g^{\nu\tau}( 2p+q')_\xi  +g^{\tau\xi}2(p+q')^\nu) \nn 
\sim {3\ g^4\ \Xi_{\la\la'}\o2(2m_Q)^6} \ep^*_{\la\eta }(\v q)\ep^*_{\la'\tau }(-\v q) (g^{\eta\xi}2(q-p)^\mu-g^{\eta\mu}( 2 p+q)^\xi)
(-g^{\nu\tau}( 2p+q')_\xi  +g^{\tau\xi}2(p+q')^\nu) \nn (\bar\Psi_{2 }(\v 0)\c_\mu \Psi_{1 }(\v 0)) (\bar\Psi_{4 }(\v 0)\c_\nu \Psi_3(\v 0))\ ,
\nonumber\eea 
where, considering that the tetra-Quarks are  in a color singlet state, we have used Equation (\ref{defi}) and few properties of the $SU(3)$ matrices.
Then shortening our notation, we get \footnote{Remember that we disregard the part of the effective Hamiltonian which   contains Quark or anti-Quark creation  and gluon destruction operators.}
\bea&&  H_a\sim=- {3\o2}{ g^4\ \Xi_{\la\la'} \o  (2 m_Q)^4}\tilde B_{s_2}(0)\tilde A_{s_1}(0)\nn \tilde B_{s_4}(0)\tilde A_{s_3}(0)\[ 
5(\s_2\ \v\s\.\v\ep^*)_{s_2 s_1}(\s_2\ \v\s\.\v\ep^{'*})_{s_4 s_3}
+4(\v \ep^* \.\v \ep^{'*})(\s_2\ \v\s\.\v v)_{s_2 s_1}(\s_2\ \v\s\.\v v)_{s_4 s_3}
\]\ .\nonumber\eea

The same kind of calculation is applied to the remaining nine diagrams.

Then we  collect the ten operators into a single effective transition Hamiltonian
 getting
\bea&& H_{eff}|_{dist} \sim -{  g^4 \Xi_{\la\la'} \o \ (2m_Q)^4} \tilde B_{s_2}(0)\tilde A_{s_1}(0)\tilde B_{s_4}(0)\tilde A_{s_3}(0)
 \[{101\o6}(\s_2\v\s\.\ep^*)_{s_2s_1})(\s_2\v\s\.\ep^{'*})_{s_4s_3})\acca+{29\o4}(\ep^*\.\ep^{'*})(\s_2\ \v \s\.\v v)_{s_2 s_1}(\s_2\ \v\s\.\v v)_{s_4 s_3}-{7\o3}\ep^*\.\ep^{'*} (\s_2 \s_i)_{s_2 s_1}(\s_2  \s_i)_{s_4 s_3}\acca -{1\o12}(\ep^*\.\ep^{'*})(\s_2 )_{s_2 s_1}(\s_2 )_{s_4 s_3}
\]
\ ,\nonumber\eea
and we  can compute the transition amplitudes in the ${\cal T}$ rest frame which are given by
\be  \d(\v k+\v k')T_{\la,\a,\v k,\ \la',\b,-\v k,\ {\cal T}_{J,J_z},\v 0}\equiv<0| a _{\la \a}(\v k) a _{\la' \b}( \v k') H_{eff}|{\cal T}_{J,J_z},\v 0> \ .\ee
Following the above introduced procedure and setting
$ \v k= 2 m_Q  \v v\ ,$ we have the creation operators in the gluon Fock space
\bea&&<0|H_{eff}|{\cal T}_{0},\v 0>_{NR}\simeq {101\o18}{  g^4\ \Xi_{\la\la'} \o \ (2m_Q)^4}\Psi_{\cal T}(\v r_i=0,\v P=0)(\ep\.\ep')\nn={101\  \a_S^2\ \Xi_{\la\la'} \o2^{5\o2}\ 3^2\ \pi^{7\o4}\ m_Q^4 \ d^3\ \d^{3\o2}}\d_{\la\la'} \ ,\nn 
<0|  H_{eff}|{\cal T}_{1,1},\v 0>_{NR}\simeq 0\ ,\nn
<0|  H_{eff}|{\cal T}_{2,2},\v 0>_{NR}\simeq {  g^4\ \Xi_{\la\la'} \o\sqrt3\  \ (2m_Q)^4}\Psi_{\cal T}(\v r_i=0,\v P=0)[{101\o6}\ \ep^*_+\ep^{'*}_++{29\o4} v_+^2(\ep^*\.\ep^{'*})]\nn={  \a_S^2\ \Xi_{\la\la'} \o2^{7\o2}\ 3^{3\o2}\ \pi^{7\o4}\ m_Q^4 \ d^3\ \d^{3\o2}}[202\ \ep^*_+\ep^{'*}_++87\ v_+^2(\ep^*\.\ep^{'*})]\ ,\label{2ga}\eea where for a generic vector $V$ we set $ V^*_+\equiv V^*_x-iV^*_y$ in the laboratory frame.

Notice that the transition matrix element ${\cal T}_{1,1}\leftrightarrow 2g$  vanishes due to the charge conjugation invariance.

Then we have
 $$ T_{\zeta,\a,\v k,\ \zeta',\b,-\v k,\ {\cal T}_{0,\v 0}}\simeq{101\  \a_S^2\  \o2^{7\o2}\ 3^2\ \pi^{7\o4}\ m_Q^5 \ d^3\ \d^{3\o2}} \d^{\a\b}\d_{\zeta\zeta'}\ ,$$
 and
$$ T_{\zeta,\a,\v k,\ \zeta',\b,-\v k,\ {\cal T}_{2,2,\v 0}}\simeq{  \a_S^2\ \o2^{9\o2}\ 3^{3\o2}\ \pi^{7\o4}\ m_Q^5 \ d^3\ \d^{3\o2}}\d^{\a\b}[202\ \ep^*_{\zeta +}(\v k)\ep^{'*}_{\zeta'  +}(-\v k)+87\ v_+^2(\v\ep^*_{\zeta}(\v k)\.\v \ep^{'*}_{\zeta'  }(-\v k)\ep')]\ .$$
Computing the Drell-Yan production cross section we need the average of the values of $ |T|^2$ over the helicity and color states of the gluons and, for ${\cal T}_2$, over the gluon momentum directions. This average is trivial for ${\cal T}_0$ which is isotrope, and  for which  we have 
\be \overline{|T_{h,\a,\v v,\ h',\b,-\v v,\ {\cal T}_0,\v 0}|^2} \simeq   10^{-3} {\a_S^4\o \ m_Q^{10} \ d^6\d^{3 }}\ ,\label{tm0}\ee
On the contrary   averaging $|{\cal T}_2|^2$  over the gluon momentum directions we must compute the dependence of the gluon polarization vectors and of the gluon momenta in the laboratory frame. For this we have to recall the rotation matrix relating the  reference system identified by the gluon momentum and polarization to the laboratory frame.

As a matter of fact we have defined in Appendix C, Equation (\ref{pol}), the gluon polarization vectors in a Cartesian reference  frame whose $z$ axis is parallel to the vector $\v k$, the gluon momentum, and whose $x$ axis is parallel to the vector $(\v\ep_{\zeta= +}(\v k)+\v\ep_{\zeta= -}(\v k))/\sqrt2$. However the computed transition amplitudes depend on the vector components  in the laboratory frame.
 In order to complete our calculation we have to introduce the Euler angles which parametrize the rotation matrix between the two frames. 
This matrix is given by
\be {\bf R}=\(\matrix{\cos\a\cos\b\cos\c-\sin\a\sin\c &-\cos\a\cos\b\sin\c-\sin\a\cos\c&\sin\b\cos\a \cr \sin\a\cos\b\cos\c+\cos\a\sin\c&-\sin\a\cos\b\sin\c+\cos\a\cos\c &\sin\b\sin \a \cr - \sin\b\cos\c & + \sin\b\sin\c & \cos\b\cr }\)\ ,\ee where $0<\a<2\pi\ ,0<\c<2\pi $ and $0<\b<\pi\ .$
 Then the complex conjugate  $+$ components of the polarization vector   with helicity $\zeta$  in the laboratory frame are
 $$\ep^*_{+, \zeta}(\pm\v k)\equiv \ep^* _{x, -\zeta}(\pm\v k)+i\ep^*_{y,- \zeta}(\pm\v k) =e^{  i(\a \mp\zeta\c)}{\cos\b\pm\zeta\o\sqrt2}\ .$$
Therefore   we can compute the average over the gluon directions of the square transition amplitude
\bea&&{ 1\o8\pi^2}{\sum_{\zeta, \zeta'}\o 4}\int_0^{2\pi}d\a d\c\int_{-1}^1 d\cos\b |202\ \ep^*_{\zeta +}(\v k) \ep^*_{\zeta' +}(-\v k)+87\ \d_{\zeta\zeta'} v_+^2 |^2 \simeq 2.09\ 10^4\ \d_{\zeta\zeta'}\ , \nonumber\eea
 and hence 
\be \overline{|T_{h,\a,\v v,\ h',\b,-\v v,\ {\cal T}_2,2,\v 0}|^2}\simeq 2.65\ 10^{-3} {\a_S^4\o \ m_Q^{10} \ d^6\d^{3 }} \ .\label{tm2}\ee Inserting into Eq. (\ref{sprod}) the results given in Eq.'s (\ref{tm0}) and (\ref{tm2})
 and taking into account the spin degeneracy of ${\cal T}_2$ one gets a production cross section which is roughly a factor $17$ greater than that of ${\cal T}_0$.

\subsection{The transition ${\cal T}\leftrightarrow q\bar q$.}
\bigskip

 Now we come to the light parton quarks  noting that at the same order one might consider the exclusive decay of the tetra-Quark into a $Q\bar Q$ pair. The rate of this decay which is of the fourth order in $\a_S$ is determined by all the diagrams in Figure 3. It is not difficult to verify that the decay rate is of the order of $\a_S^4/(m_Q^8\ \d^3\ d^6)$, that is of few electron-Volt and hence negligible.

In fact we are interested in the transition amplitude ${\cal T}\to q\bar q$ for light quarks which determines the parton quark contribution to the production cross section of the ${\cal T}$. This is interesting even if we know that the corresponding luminosity is by almost two orders of magnitude smaller than that of two gluons. 

It is apparent that the effective Hamiltonian only corresponds to diagram (h).
We shall systematically follow the simplifications and notations used in the previous Section replacing the gluons by the light quarks.
We denote by $a$ and $b$ the destruction operators of the light quarks and anti-quarks and neglect their masses.

\begin{figure}[h]
\centering
\subfigure[]{\includegraphics[width=0.1\textwidth]{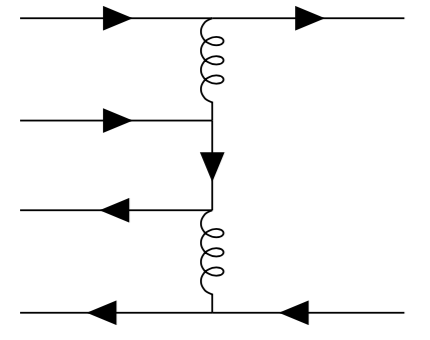}} 
\subfigure[]{\includegraphics[width=0.1\textwidth]{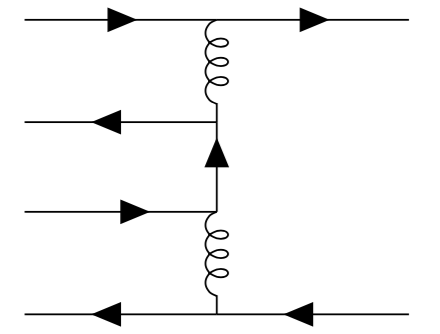}} 
\subfigure[]{\includegraphics[width=0.1\textwidth]{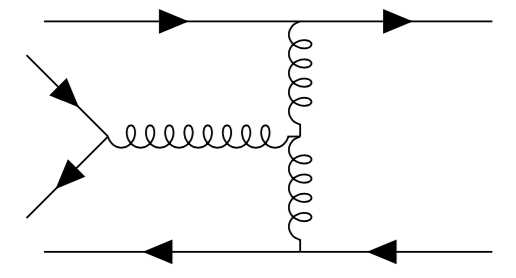}}
\subfigure[]{\includegraphics[width=0.1\textwidth]{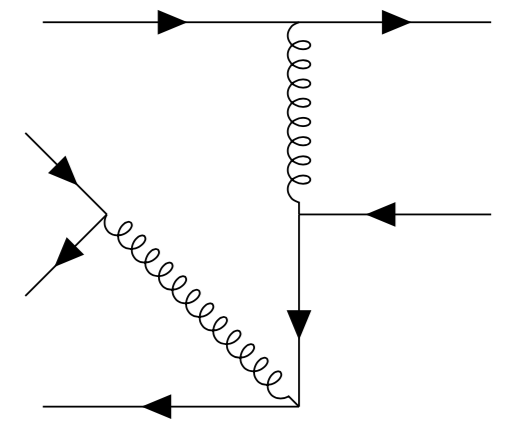}}
\subfigure[]{\includegraphics[width=0.1\textwidth]{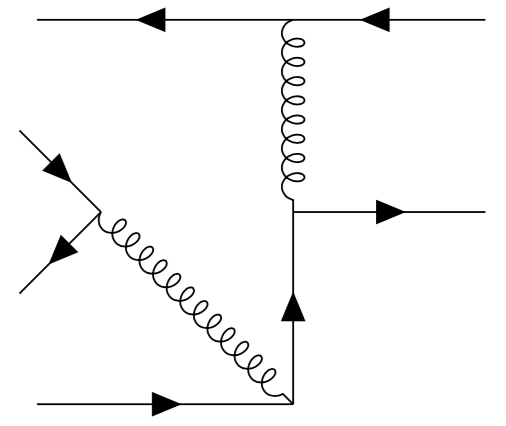}}
\subfigure[]{\includegraphics[width=0.1\textwidth]{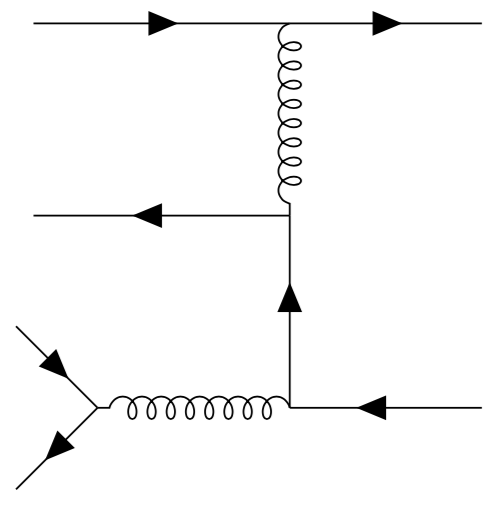}}
\subfigure[]{\includegraphics[width=0.1\textwidth]{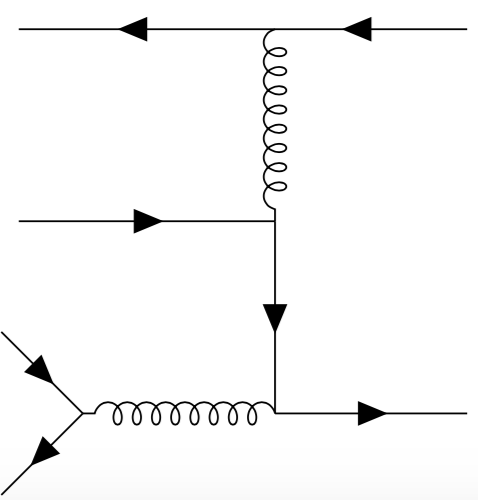}}
\subfigure[]{\includegraphics[width=0.1\textwidth]{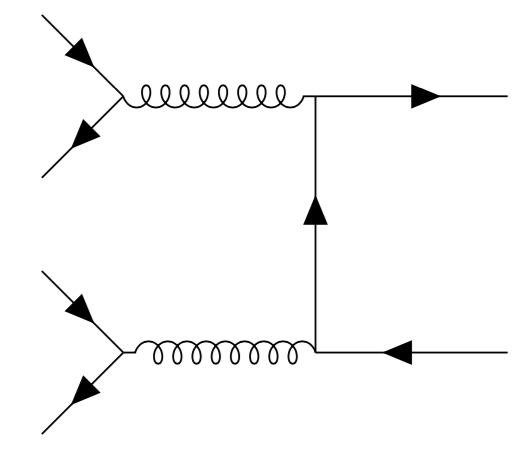}}
\caption{The transition ${\cal T}\leftrightarrow q\bar q$. }
\label{fig:HeavyLight}
\end{figure}
Then we have the effective Hamiltonian
\bea&&H_{esc}\simeq -{g^4\o\ (2  m_Q)^5}
\tilde B_{s_4,c_4}(\v 0)\tilde A_{s_3,c_3}(\v 0)\tilde B_{s_2,c_2}(\v 0)\tilde A_{s_1,c_1}(\v 0)
(\bar U_{s_4}\c_2 \c^\nu U_{s_3})(\bar U_{s_2}\c_2 \c^\mu U_{s_1})\nn  t^\a_{c_4 c_3}t^\b_{c_2 c_1} (t^\a t^\b)_{lm} \int {d\v q\o  2 q }a^\dag_{ h ,m}(\v q)b^\dag_{ h',l}(-\v q)(\bar u_{h }( \v q) \c_\mu \vs\c_\nu 
v_{h'}(-\v q))\nn=
 {\pi^2\ \a_S^2\  \o 3\ m_Q^5\ } 
\tilde B_{s_4,c_4}(\v 0)\tilde A_{s_3,c_3}(\v 0)\tilde B_{s_2,c_2}(\v 0)\tilde A_{s_1,c_1}(\v 0)\ 
 ( \s_2\s_i)_{s_2s_1}( \s_2\s_j)_{s_4s_3}\nn \int {d\v q\o  2 q }a^\dag_{ h ,m}(\v q)b^\dag_{ h',m}(-\v q)(\bar u_{h }( \v q) \c_i \vs\c_j v_{h'}(-\v q))
\ .\label{2qq}\eea

The corresponding mean square transition amplitudes are  \bea&& \overline{|T_{h,m,\v v,\ h', l,-\v v,\ {\cal T}_0,\v 0}|^2}\simeq{ \a_S^4\o3^5\ 2^7\pi^{7\o2}\ d^6\d^3\ m_Q^{12}}\int {d\v v\o4 \pi}\ m_Q^2Tr[(\c_0+\vs)\vs(\c_0-\vs)\vs]=0\nn \overline{|T_{h,m,\v v,\ h', l,-\v v,\ {\cal T}_2,2,\v 0}|^2}\simeq { \a_S^4\o3^4\ 2^5\pi^{7\o2}\ d^6\d^3\ m_Q^{12}}\int {d\v v\o4 \pi}\ m_Q^2 Tr[(\c_0+\vs) ( \vs +i(v_x\c_y+v_y\c_x))\nn (\c_0-\vs)( \vs -i(v_x\c_y+v_y\c_x))]={ \a_S^4\o3^4\ 10\ \pi^{7\o2}\ d^6\d^3\ m_Q^{10}}\simeq 1.2\ 10^{-3} {\a_S^4\o \ m_Q^{10} \ d^6\d^{3 }} \ .\label{tm2q}\eea
Thus, the mean square  transition amplitude between ${\cal T}_2\to q\bar q $ 
  is about half   that in two gluons. Therefore, since the gluon-gluon luminosity is about two orders of magnitude greater than the quark-anti-quark one, this last contribution can be  disregarded.

\section*{Acknowledgements}
\thispagestyle{empty}

The Author is in debt of gratitude to E. Santopinto for constant help and encouragement, to S. Frixione, S. Marzani and G. Ridolfi for their crucial help on the parton-parton luminosity calculations. He is also in debt to F. Parodi for information on detection efficiencies at the LHC.

\end{document}